\documentclass[aps,pra,preprint,
superscriptaddress, preprintnumbers,
titlepage,floatfix, longbibliography, nobibnotes
]{revtex4-2}

\usepackage{amsmath, amsfonts, amssymb, bm, dcolumn}
\usepackage{dsfont}
\usepackage{bm}
\usepackage{graphicx}
\usepackage[normalem]{ulem}

\usepackage[
  colorlinks=true,
  linkcolor=blue,
  citecolor=red,
  urlcolor=magenta,
  hypertexnames=false
]{hyperref}

\begin{document}
\preprint{}

\title{Discovery of spontaneous mesoscopic strain waves in nematic domains using
dark-field X-ray microscopy}

\author{Kaan Alp \surname{Yay}}
\email[]{kaanalpyay@stanford.edu}
\affiliation{Geballe Laboratory for Advanced Materials, Stanford University, Stanford, CA, USA}
\affiliation{Stanford Institute for Materials and Energy Sciences, SLAC National Accelerator Laboratory, Menlo Park, CA, USA}
\affiliation{Department of Physics, Stanford University, Stanford, CA, USA}

\author{W. Joe Meese}
\affiliation{Department of Physics, The Grainger College of Engineering,
University of Illinois Urbana-Champaign, Urbana, IL, USA}
\affiliation{Anthony J. Leggett Institute for Condensed Matter Theory, The Grainger College of Engineering,
University of Illinois Urbana-Champaign, Urbana, IL, USA}

\author{Elliot Kisiel}
\author{Matthew J. Krogstad}
\affiliation{X-ray Science Division, Advanced Photon Source, Argonne National Laboratory, Argonne, IL, USA}

\author{Anisha G. Singh}
\affiliation{Geballe Laboratory for Advanced Materials, Stanford University, Stanford, CA, USA}
\affiliation{Stanford Institute for Materials and Energy Sciences, SLAC National Accelerator Laboratory, Menlo Park, CA, USA}
\affiliation{Department of Applied Physics, Stanford University, Stanford, CA, USA}

\author{Rafael M. Fernandes}
\affiliation{Department of Physics, The Grainger College of Engineering,
University of Illinois Urbana-Champaign, Urbana, IL, USA}
\affiliation{Anthony J. Leggett Institute for Condensed Matter Theory, The Grainger College of Engineering,
University of Illinois Urbana-Champaign, Urbana, IL, USA}

\author{Zahir Islam}
\email{zahir@anl.gov}
\affiliation{X-ray Science Division, Advanced Photon Source, Argonne National Laboratory, Argonne, IL, USA}

\author{Ian R. Fisher}
\email{irfisher@stanford.edu}
\affiliation{Geballe Laboratory for Advanced Materials, Stanford University, Stanford, CA, USA}
\affiliation{Stanford Institute for Materials and Energy Sciences, SLAC National Accelerator Laboratory, Menlo Park, CA, USA}
\affiliation{Department of Applied Physics, Stanford University, Stanford, CA, USA}

\date{July 31, 2025}

\begin{abstract}
\textbf{Electronic nematic order is a correlated phase of matter in which low-energy electronic states spontaneously break a discrete rotational symmetry of a crystal lattice. Bilinear coupling between the electronic nematic and strains of the same symmetry yields a single pseudoproper ferroelastic phase transition at which both the nematic and lattice strain onset concurrently. To minimize elastic energy, the crystal forms structural twin domains, each with a distinct orientation of the nematic director (i.e. each with a specific sign of the induced shear strain). While the effects of externally induced strains on these domains are well established, the intrinsic behavior of spontaneous strain fields within individual domains has been hitherto unexplored, largely due to the lack of appropriate experimental tools. Here, we report the discovery of spontaneous mesoscopic strain waves within individual nematic domains of an underdoped iron-based superconductor, observed using dark-field X-ray microscopy (DFXM). This technique combines high spatial and reciprocal-space resolution with full-field, bulk-sensitive imaging, enabling direct visualization of subdomain strain modulations emerging concurrently with the onset of nematic order. The elastic compatibility relations that govern inhomogeneous strains in continuous solids provide a natural mechanism for the emergent strain waves that we observe. Our findings reveal a broadly relevant form of strain self-organization and position DFXM as a powerful tool for probing the local interplay between lattice strain and electronic order.}
\end{abstract}

\maketitle

The spontaneous emergence of symmetry-breaking strain is a hallmark of ferroelastic phase transitions in solids~\cite{Salje2012FerroelasticMaterials}. For temperatures below the critical temperature, the crystal lattice can deform in at least two (depending on the crystal symmetry) equivalent, but distinct orientations related to each other by the broken symmetry. To reduce the overall strain energy, the crystal often forms structural twin domains in the broken-symmetry phase---mesoscale regions with differing orientations of the spontaneous lattice deformation~\cite{Tagantsev2010DomainsFilms}. These domains are then separated by well-defined interfaces called domain walls. While the symmetry properties of structural domains are well established~\cite{Tagantsev2010DomainsFilms, Salje1993PhaseCrystals}, how internal local strain fields behave at the mesoscopic scale inside bulk domains is largely unexplored, especially in quantum materials for which the structural transition is driven by electronic degrees of freedom (e.g., in electronic nematic systems \cite{Kivelson98,Fradkin_review,fernandesWhatDrivesNematic2014}). In addition to the spontaneous strain of the broken-symmetry phase, extrinsic strain fields readily form around common lattice imperfections such as dislocations~\cite{Hytch2003MeasurementMicroscopy} and internal boundaries such as domain walls~\cite{Hytch1998QuantitativeMicrographs}, where they can reach a magnitude of $5 \times 10^{-4}$ and extend for several microns~\cite{Simons2018Long-rangeFerroelectrics}. Long-range inhomogeneous strain fields at these magnitudes can affect the local electronic properties of a variety of bulk quantum materials. Notably, anisotropic strains can lead to strong resistive anisotropies~\cite{Chu2010In-PlaneSuperconductor, Chu2012DivergentSuperconductor}, suppress~\cite{Malinowski2020SuppressionPoint} or enhance~\cite{Steppke2017StrongPressure} superconductivity, reorder the charge density wave vector~\cite{Straquadine2022EvidenceMeasurements}, and drive the band structure across a topological transition~\cite{Sunko2019DirectSr2RuO4,Mutch2019EvidenceZrTe5} for several distinct strongly-correlated and topological systems. Thus, it is of critical importance to obtain a quantitative understanding of local strain variations that must occur at/near domain boundaries. The present study focuses on nematic domains, but the ideas that we advance are more broadly relevant, applying to any system for which spontaneous domain formation results in local strain gradients.  

Methods commonly used in condensed-matter physics research to measure strain, however, either do not have the spatial resolution to detect the strain variation at the micrometer scale or are unable to sample a large enough volume in the interior of a bulk crystal. On the one hand, methods such as strain gauges~\cite{Chu2012DivergentSuperconductor, Kuo2016UbiquitousSuperconductors},  displacement sensors~\cite{Hicks2014Piezoelectric-basedTuning}, and conventional X-ray diffraction~\cite{Sanchez2021TheDiffraction, Singh2024EmergentMaterial} only give an average picture of the bulk strain at the millimeter scale. On the other, high spatial resolution methods such as TEM require sample thicknesses below 100 nm and are therefore unable to probe strain fields within bulk samples~\cite{Tagantsev2010DomainsFilms}. Recent application of focused ion beam milling on strongly-correlated metals has enabled researchers to tailor the geometry of samples and thus impose specific mechanical boundary conditions at micrometer length scales~\cite{Bachmann2019SpatialCeIrIn5, Moll2018FocusedMatter}. Achieved strain magnitudes and orientations in the sample are then predicted by finite-element analysis (FEA) simulations. Although the predicted strain values have been in accordance with transport and magnetization measurements in ultraclean samples~\cite{Bachmann2019SpatialCeIrIn5}, FEA simulations do not take into account local chemical strain due to stoichiometric deviations~\cite{Freedman2009ElasticStrain} or internal structure such as ferroelastic domain walls, both of which are common in materials of high interest, including doped YBCO~\cite{Schmahl1989TwinYBa2Cu1-xCox3O7-, Tagantsev2010DomainsFilms} and BaFe$_2$As$_2$ \cite{Blomberg2012EffectBaFe2As2, Prozorov2009IntrinsicBaFe1-xCox2As2}. Thus, we lack mesoscopic studies of internal strain fields in quantum materials.

To address this challenge, we employ dark-field X-ray microscopy (DFXM), a novel full-field imaging technique~\cite{Simons2015Dark-fieldCharacterization}. As in dark-field electron microscopy, the idea of DFXM is to insert an objective lens into the path of a particular Bragg peak to obtain a magnified real-space image of the regions of the crystal that are diffracting toward that Bragg peak. Using X-rays instead of electrons, however, enables DFXM to non-destructively image the interior regions of samples of thickness exceeding 100 microns. By positioning the lens downstream of the sample, and thanks to the high spatial and angular resolution of modern X-ray lenses~\cite{Qiao2020AOptics}, one can resolve strains of $\sim$$10^{-5}$ in a field of view of 30-100 $\mu$m with a spatial resolution of 50-100 nm. DFXM has been applied successfully to visualize the orientation and strain variations due to crystal heterogeneities in a variety of industrial and functional materials such as aluminum \cite{Simons2015Dark-fieldCharacterization, Dresselhaus-Marais2021InMelting} and polycrystalline BaTiO$_3$~\cite{Simons2018Long-rangeFerroelectrics}. More recently, the extension of DFXM techniques to cryogenic temperatures~\cite{Plumb2023DarkNaMnO2} has been used to image the spatial separation of two distinct charge-density wave orders and the related structural twinning in the Kagome superconductor CsV$_3$Sb$_5$~\cite{Plumb2024Phase-separatedCsV3Sb5}.

Here, we report the direct observation of coherent micron-scale strain waves within single nematic/ferroelastic domains in the low-temperature orthorhombic phase of the correlated iron pnictide Ba(Fe$_{0.98}$Cu$_{0.02}$)$_2$As$_2$ using cryogenic DFXM. The strain waves have an amplitude of $10^{-5}$ with a wavelength of $\sim$2-3 $\mu$m and run parallel to the orthorhombic twin domain walls. As the twin domain walls disappear in the high-temperature tetragonal phase, so do the strain waves. We propose that the coherent strain waves emerge to satisfy the local elasticity compatibility relations of inhomogeneous strain fields and the mechanical boundary conditions imposed on the domain by its boundaries. Our result highlights that strain must be treated as a spatially varying local field rather than a uniform field, since mesoscopic strain variations must obey additional geometrical constraints. 

\section*{\texorpdfstring{\hspace*{-\parindent}Nematic domains in iron pnictides}{Nematic domains in iron pnictides}}
\label{sec:nematic-domains}

Ba(Fe$_{1-x}$Cu$_x$)$_2$As$_2$ is an iron pnictide in the same family as the archetypal iron-based superconductor Ba(Fe$_{1-x}$Co$_x$)$_2$As$_2$~\cite{Chu2009DeterminationBaFe1-xCox2As2, Paglione2010High-temperatureMaterials, Fernandes2022IronSuperconductivity}. The parent compound BaFe$_2$As$_2$ undergoes two phase transitions near 134 K: first a ferroelastic structural phase transition at $T_S$, and then an antiferromagnetic transition at $T_N$. Similar to Co doping of the Fe sites, Cu doping suppresses both of these transitions to lower temperatures~\cite{Ni2010TemperatureCrystals}. In contrast to the extensive superconducting dome induced by Co doping, however, Cu doping is reported to induce superconductivity only in the vicinity of a doping level of $x=4.4\%$ with an optimal $T_c = 2.1$ K~\cite{Ni2010TemperatureCrystals}, presumably due to the pair-breaking effects of Cu \cite{Fernandes2012}. The sample we used in our measurements was a single crystal of Ba(Fe$_{0.98}$Cu$_{0.02}$)$_2$As$_2$ (Cu-Ba122) with two well-defined phase transitions at $T_S = 94.1 \pm 1.5$ K and $T_N = 85.7 \pm 1.7$ K, and no observable superconducting transition above 1.8 K. 

\begin{figure}[p]
    \centering
    \includegraphics[width=\linewidth]{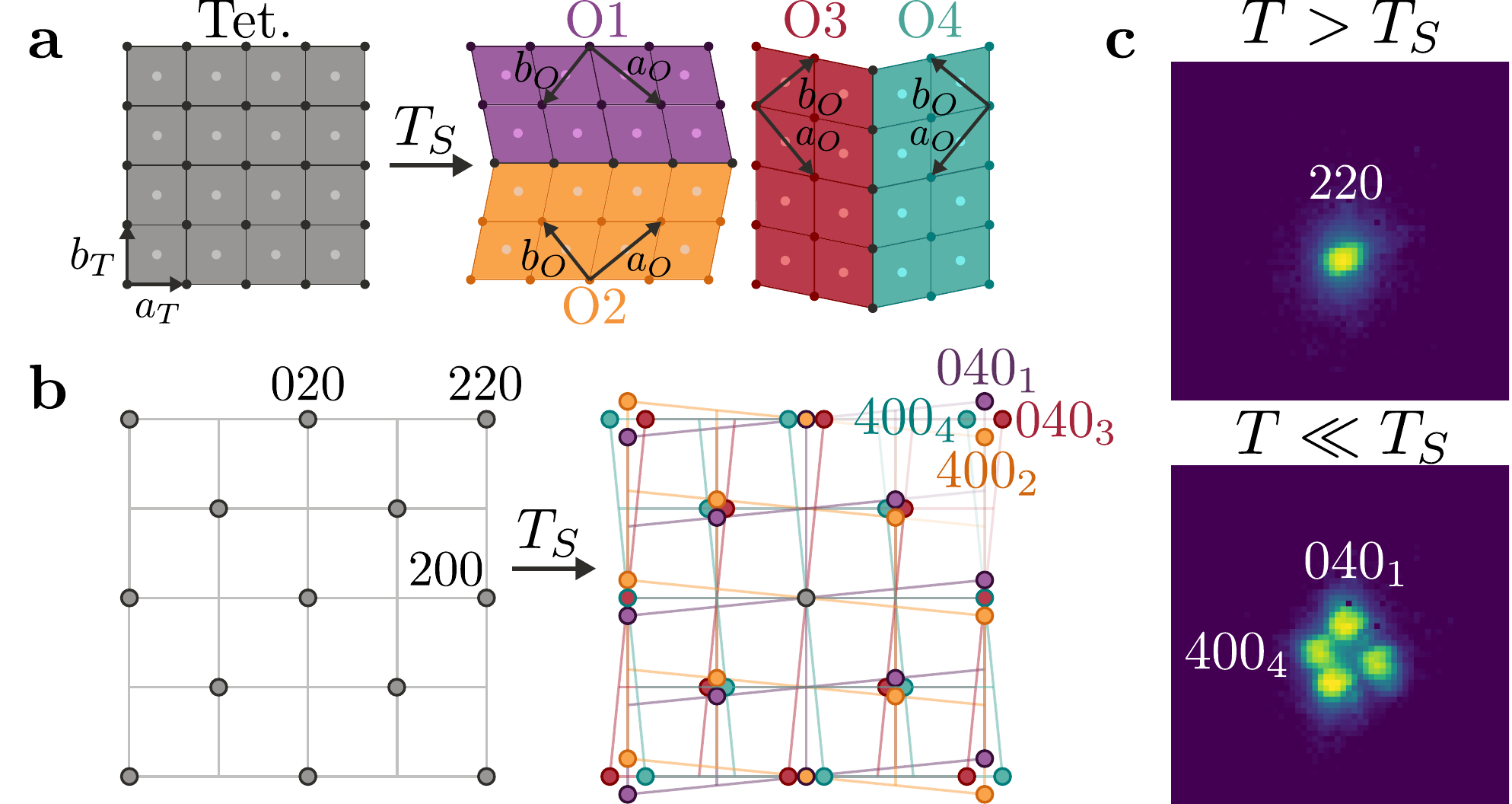}
    \caption{\textbf{Signatures of the nematic structural phase transition in reciprocal space} 
    \newline
    \textbf{a} (left) Body-centered tetragonal crystal structure of Cu-Ba122 at high temperature viewed from the $c$ axis. (right) Four possible face-centered orthorhombic domains formed as a result of the nematic structural phase transition at temperature $T_S \simeq 94$ K. Orthorhombic primitive lattice vectors $a_O$ and $b_O$ are rotated by $\simeq 45^\circ$ with respect to tetragonal $a_T$ and $b_T$, where we denote $a_O$ as the long axis. \textbf{b} Reciprocal space picture of the structural phase transition. $220$ reflection of tetragonal phase splits into a quartet due to formation of twin domains. Subscripts of the quartet reflections correspond to the different domains in panel~\textbf{a}. \textbf{c} (above) 220 reflection peak of Cu-Ba122 sample in tetragonal phase at $T = 100$~K as measured by high-energy X-ray diffraction. The incident beam is to the bottom left of the image with respect to the peak. (below) Fully split 220 peak at base temperature $T = 10$~K where the peaks selected for DFXM measurements are labeled. The distortion of the quartet from the schematic in \textbf{b} is presumably due to the built-in strain and domain distribution as the sample is cooled through $T_S$ \cite{Tanatar2010Uniaxial-strainStudy, Blomberg2012EffectBaFe2As2}.
    }
    \label{fig:fig1}
\end{figure}

At room temperature, Cu-Ba122 has a tetragonal unit cell with space group $I4/mmm$. The ferroelastic phase transition at $T_S$ drives the crystal into an orthorhombic phase with space group $Fmmm$, in which the unit cell of the tetragonal phase develops a spontaneous symmetry-breaking strain along the $[110]_T$ direction—referred to as $\varepsilon_{xy}$ or $\varepsilon_{B_{2g}}$ with respect to the tetragonal unit cell. This ferroelastic phase transition is a pseudoproper one: the critical order parameter is the electronic nematicity of $B_{2g}$ symmetry which also induces a spontaneous strain of the same symmetry through a bilinear nematoelastic coupling~\cite{Chu2012DivergentSuperconductor, Kuo2013MeasurementBaFe0.975Co0.0252As2}. The spontaneous strain can have a positive value in the $[110]_T$ or $[1\bar{1}0]_T$ direction, defining two orientational domain states. These two domain states can form twins along either of the two permissible domain wall directions of $[100]_T$ and $[010]_T$, leading to a total of four distinct suborientational orthorhombic domain types, as depicted in Fig.~\ref{fig:fig1}\textbf{a}. Furthermore, the unit cell in the orthorhombic phase is doubled and rotated by $45^\circ \pm \alpha$ with respect to that in the tetragonal phase, where $\alpha = \varepsilon_{xy} \approx \frac{a_O-b_O}{a_O+b_O}$. We use the convention that $a_O$ ($b_O$) refers to the long (short) in-plane axis for each domain in the orthorhombic phase. The clapping angle $\alpha$ arises because the component domains of each twin rotate toward their corresponding twin boundary by $\alpha$ to stay in physical contact despite the orthorhombic distortion~\cite{Tagantsev2010DomainsFilms}.

In real space, several surface- and bulk-sensitive methods have been applied to image the spatial dimensions of the orthorhombic domains in iron pnictides. Polarized light microscopy measurements have detected stripe-like domains elongated along the twin boundaries with domain widths of $5$-$100$ $\mu$m in a variety of undoped, Co-doped and P-doped BaFe$_2$As$_2$ samples~\cite{Tanatar2009DirectBa, Prozorov2009IntrinsicBaFe1-xCox2As2, Thewalt2018ImagingBaFe2AsP2, Yang2020NematicGas}. Moreover, bulk-sensitive measurements such as scanning SQUID microscopy~\cite{Kalisky2010StripesBoundaries} and quantum gas magnetometry~\cite{Yang2020NematicGas} confirmed the dimensions observed in surface measurements and showed that the twin domains extend along the c axis into the bulk of the crystal. Despite the success of these techniques in detecting the shape and size of the twin domains in the orthorhombic phase, they all suffer from the shortcoming that their spatial resolution is at best several microns and cannot detect variations at that length scale. Moreover, none of the above techniques can directly probe bulk strain fields within the domains. While it is true that the angle of rotation of linearly polarized light upon reflection from a domain is proportional to the orthorhombic distortion~\cite{Yang2020NematicGas}, the skin depth of a few tens of nanometers in iron pnictides~\cite{Stojchevska2012DopingFluctuations} renders optical microscopy measurements unable to probe the full depth of bulk samples with thicknesses of tens of microns.

In reciprocal space, the formation of ferroelastic twins below $T_S$ results in the splitting of the Bragg peaks of the tetragonal unit cell. Specifically, the $220_T$ reflection splits into four peaks with each peak corresponding to a different orthorhombic domain type, as demonstrated in Fig.~\ref{fig:fig1}\textbf{b}. Moreover, the splitting becomes larger at lower temperatures due to the growth of the degree of orthorhombicity and clapping angle as temperature is lowered. This peak splitting and the formation of twin domains were first observed in X-ray diffraction studies on bulk samples of undoped and several Co-doping values of BaFe$_2$As$_2$~\cite{Tanatar2009DirectBa, Prozorov2009IntrinsicBaFe1-xCox2As2}. 

We confirmed this phenomenon in Cu-Ba122 in our preliminary reciprocal space measurement: using high-energy X-ray diffraction with a beam spot of 500 x 500 $\mu$m$^2$ on a bulk single-crystal of Cu-Ba122, we observed a well-defined $220_T$ reflection of the tetragonal lattice at high temperatures. The $220_T$ reflection then splits at $T_S \simeq 94$ K into four peaks as shown in Fig.~\ref{fig:fig1}\textbf{c}, indicating that the illuminated volume of the sample does indeed contain all four types of domains in the orthorhombic phase. The slight distortion of the quartet peaks from a square can be explained by the built-in strain in the illuminated volume and the resulting domain distribution as the sample is cooled below $T_S$. Such distortions of the quartet pattern have been shown in undoped BaFe$_2$As$_2$ to be caused by domain realignment when external strain is applied on the sample in the $[110]_T$ direction~\cite{Tanatar2010Uniaxial-strainStudy, Blomberg2012EffectBaFe2As2}. Since the illuminated volume in our measurement was free-standing, we ascribe the distortion in our measurement to built-in strain fields within the sample due to growth conditions.

\section*{\texorpdfstring{\hspace*{-\parindent}Measurement of strain waves using DFXM}{Measurement of strain waves using DFXM}}\label{sec:dfxm}

\begin{figure}[p]
    \centering
    \includegraphics[width=\linewidth]{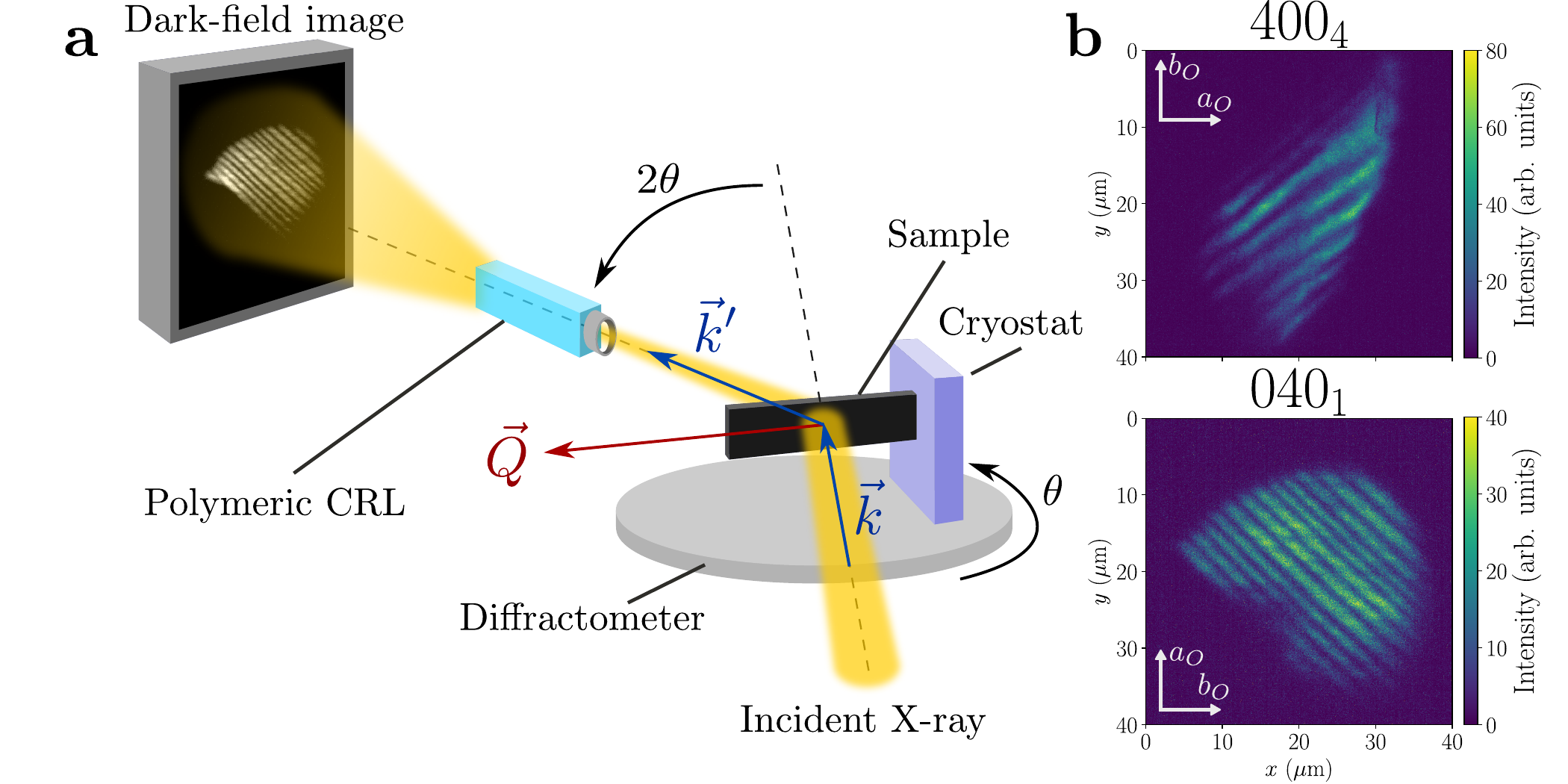}
    \caption{\textbf{Imaging periodic scattering intensity modulations within individual nematic domains using dark-field X-ray microscopy}
    \newline
    \textbf{a} Dark-field X-ray microscopy (DFXM) experimental setup. Cu-Ba122 sample is mounted from one end on a cryostat installed on a horizontal diffractometer. A free-standing volume of the sample is illuminated by the incident X-ray beam in transmission geometry. A polymeric compound refractive lens (CRL) installed on the $2\theta$ arm acts as an objective to image and magnify the scattered light emanating from the illuminated volume corresponding to a selected diffraction peak. The sample ($\theta$) and the CRL ($2\theta$) are rotated to capture the image of the desired Bragg peak (with scattering vector $\vec{Q}$) on the detector and also to scan the real-space distribution of axial strain and lattice orientation with respect to $\Vec{Q}$. $\vec{k}$ and $\vec{k}'$ correspond to the wave vectors of the incident and scattered radiation, respectively. 
    \textbf{b} Real-space microscope images of the $400_4$ and $040_1$ peaks labeled in Fig.~\ref{fig:fig1}c. Each image depicts a single domain with a mesoscopic spatial diffraction intensity modulation as explained in the main text. Above image was collected at $T = 3$~K and the below one at $T = 60$~K. The curved outer perimeter of the images are due to a pinhole at the entrance of the objective lens.
    }
    \label{fig:fig2}
\end{figure}

Dark-field X-ray microscopy is a full-field microscopy technique that enables the investigation of local spatial distribution of reciprocal space with the insertion of an X-ray objective lens in the path of the diffracted light~\cite{Simons2015Dark-fieldCharacterization, Simons2016MultiscaleMicroscopy, Yildirim2020ProbingMicroscopy, Poulsen2017X-rayOptics}. Our experimental setup for DFXM measurements consisted of a custom low-vibration cryostat installed on a diffractometer, a polymeric compound refractive lens positioned on the $2\theta$-arm to image and magnify light diffracting at a particular scattering angle $2\theta$, and an area detector positioned on the image plane of the lens, as pictured in Fig.~\ref{fig:fig2}\textbf{a}~\cite{Plumb2023DarkNaMnO2, Kisiel2025High-ResolutionDevices, Kisiel2024DirectImaging}. The small numerical aperture of the lens~\cite{Qiao2020AOptics} in combination with a pinhole placed in front of the lens prevented stray diffraction and blurring. Additionally, a pinhole inserted in the back focal plane of the lens acted as a low-pass Fourier filter for off-axis diffraction. The insertion of the pinhole in the back focal plane further improved the high angular resolution of the lens thanks to its remarkably low numerical aperture.

The high angular resolution of the imaging system corresponds to a high resolution in reciprocal space and thus allows domain-selective imaging: individual Bragg peaks in the orthorhombic phase can be selected to image the real-space distribution of the corresponding domain type. During our measurement we imaged the $400_4$ and $040_1$ peaks as labeled in Fig.~\ref{fig:fig1}\textbf{c}; representative diffraction intensity images are given in Fig.~\ref{fig:fig2}\textbf{b}. The real space images of scattering intensity show periodic stripe patterns along the $[100]_T$ and $[010]_T$ (or equivalently $[110]_O$ and $[1\bar{1}0]_O$) directions. The stripes in the $400_4$ image are along $[100]_T$ and those in the $040_1$ image are along $[010]_T$, which are the directions of the corresponding domain walls of the O4 and O1 domain types respectively.

To identify the origin of the observed intensity stripes, we employed ($\theta,2\theta$) scans, which yield real-space maps of strain along the axis of the scattering vector $\vec{Q}$, also referred to as axial strain. For our sample orientation, this strain corresponds to $\varepsilon_{xy}$ with respect to the tetragonal unit cell. The high reciprocal-space resolution of the setup enabled not only the imaging of individual Bragg peaks but also of multiple distinct reciprocal-space points within a given Bragg peak. From Bragg's law, reciprocal-space points with distinct $2\theta$ values within a Bragg peak correspond to different interplanar spacing values (also known as $d$-spacing) between the parallel lattice planes  associated with the imaged Bragg peak. As a result, regions with different
$d$-spacings within a given domain were visualized. Resolving the fine $d$-spacing variations confirmed that the observed stripe-shaped intensity patterns arise from subdomain strain waves, as shown in Fig.~\ref{fig:fig3}\textbf{a}, rather than from switching between different domain types. 

The observed strain waves have an amplitude of $\sim$$10^{-5}$ about the average $d$-spacing value of the given domain, a strain oscillation two orders of magnitude smaller than the $\sim$$10^{-3}$ strain difference between the component domains of a twin, verifying that the observation is a monodomain phenomenon. Moreover, the oscillations have a wavelength at the mesoscopic scale of $\sim$$2$~$\mu$m and permeate the imaged section of the entire domain, as shown in the zoomed image of Fig.~\ref{fig:fig3}\textbf{b}.

Upon entering the tetragonal phase, a marked change happens in the strain map of the sample. As the orthorhombicity disappears, so do the separate domains, and the quartet of peaks merges back into a single $220_T$ peak. Performing a ($\theta,2\theta$) scan on this peak reveals a strain distribution that lacks the periodic wave pattern observed in the low-temperature orthorhombic phase, as depicted in Fig.~\ref{fig:fig3}\textbf{c}. We infer from these results that the strain waves observed below $T_S$ are purely a phenomenon of the low-temperature nematic phase, where the emergence of orthorhombic twin domain walls along symmetry-allowed directions also sets the directionality of the observed strain wave.

\section*{\texorpdfstring{\hspace*{-\parindent}Orientation maps and determination of the wavelength of strain waves}{Orientation maps and determination of the wavelength of strain waves}}\label{sec:mosaic}

To further understand the phenomenon of strain waves we also investigate how lattice orientation is affected by the presence of $\varepsilon_{xy}$ strain waves. DFXM not only enables us to measure the local variation of axial strain but also permits the determination of the spatial dependence of the lattice tilt or orientation. The lattice orientation by itself does not directly measure a particular strain value; however, its continuous variation corresponds to a combination of out-of-plane strain components. We measured the lattice orientation by performing a fine $\theta$ scan at a fixed scattering angle $2\theta$ in the vicinity of a selected Bragg peak. Moreover, we performed this measurement at various adjacent sites on the sample to cover a region of $\sim100$ $\mu$m $\times$ $100$ $\mu$m. The resulting orientation map reveals a gradual and approximately monotonic evolution of the lattice tilt across a large area without a marked stripe pattern, as depicted in Fig.~\ref{fig:fig4}\textbf{a}. The longer length-scale variation in orientation is likely due to a slight bend of the lattice in the imaged volume within the crystal and appears unrelated to the detected strain waves. In contrast, the intensity image corresponding to a specific lattice tilt value used in the orientation map (Fig.~\ref{fig:fig4}\textbf{b}) exhibits pronounced stripes, indicating that the observed intensity stripes occur independently of any local variations in lattice orientation. This observation further suggests that the diffraction intensity stripes are caused predominantly by $\varepsilon_{xy}$ strain waves.

With the existence of $\varepsilon_{xy}$ strain waves established, we next determine that the wavelength of these waves do not change appreciably as a function of temperature. As we have already shown that the intensity stripes are caused by strain waves, we used the wavelength of the intensity stripes as a substitute for the wavelength of the strain waves. We performed a Fourier transform (FT) on individual images taken at a given temperature and summed the absolute value of these FTs to determine an aggregate wavelength value for that temperature. The resulting two-dimensional FT figures plotting the salient wave vectors at three different temperatures are shown in Fig.~\ref{fig:fig4}\textbf{c}. The figures are plotted in a logarithmic intensity scale to distinguish the finite wave vector peaks of interest that are orders of magnitude smaller in intensity compared to the zero-wave vector peak. This large central peak is due to the positive average value of intensity in each image. The main sinusoidal satellite peaks with highest intensity are indicated by arrows in Fig.~\ref{fig:fig4}\textbf{c} and correspond to the wavelength of $\sim$$2$ $\mu$m observed in the real space images shown in previous figures. As expected, the satellite peaks of the $040_1$ and $400_4$ Bragg peaks lie on axes rotated by 90$^\circ$ with respect to each other. A numerical investigation of the wavelength corresponding to these main satellite peaks reveal that the wavelength does not change in any significant way as a function of temperature. Besides the main satellite peaks, we also observe a set of quasi-periodic subdominant peaks in the FT images, which we discuss in the Supplementary Information.

\section*{\texorpdfstring{\hspace*{-\parindent}Compatibility relations set the wave direction}{Compatibility relations set the wave direction}}\label{sec:comprels}

The phenomenon of strain waves within bulk nematic domains of an iron pnictide we have presented has not yet been observed in any other class of quantum materials to our knowledge and challenges the conventional understanding of how strain behaves mesoscopically in materials with structural distortions. This result demonstrates that spatially coherent strain fields can emerge inside structural domains, and it is essential to model strain as a local field within materials rather than a uniform value. This insight is significant, as the intuition of homogeneous strains developed at the macroscopic level does not necessarily apply to strain fields varying at the mesoscopic level. 

Unlike homogeneous strains, locally varying strain fields must obey certain constraints. From a geometrical perspective, continuous deformations of a material will displace individual unit cells from their equilibrium positions, $\boldsymbol{r}$, to a new position,  $\boldsymbol{r}^\prime = \boldsymbol{r} + \boldsymbol{u}(\boldsymbol{r})$.  Thus, the three-component displacement \textit{vector}, $\boldsymbol{u}(\boldsymbol{r})$, represents the complete description of the deformation. The six-component strain \textit{tensor}, $\varepsilon_{ij}(\boldsymbol{r})$, is over-complete, since its six components are various derivatives of the three components of $\boldsymbol{u}(\boldsymbol{r})$. The constraints on the strain tensor reduce the number of independent strain components from \textit{six} in the \textit{homogeneous} limit to only \textit{three inhomogeneous} ones. For infinitesimal strains, $2\varepsilon_{ij}(\boldsymbol{r}) = \partial_j u_i(\boldsymbol{r}) +\partial_i u_j(\boldsymbol{r})$, which results in constraints known as the \textit{Saint-Venant Compatibility Relations} (SVCR) of continuum elasticity theory \citep{kleinertGaugeFieldsSolids1989, Meese_short_paper_2025, Meese_long_paper_2025}. The impact of the SVCR on martensite and ferroelectric domains has been previously discussed \cite{Rasmussen2001,Littlewood2014}; here, we focus on its effect on nematic domains.

Consequently, for non-uniform strains, distinct irreducible representations of the strain tensor are coupled to each other despite being orthogonal in symmetry. In systems with $\text{D}_{4h}$  point group symmetry undergoing planar deformations, the three nonzero strain components---the dilatation $\varepsilon_{A_{1g}} \equiv \varepsilon_{xx} + \varepsilon_{yy}$, the deviatoric $\varepsilon_{B_{1g}} \equiv \varepsilon_{xx} - \varepsilon_{yy}$, and the shear $\varepsilon_{B_{2g}} \equiv 2\varepsilon_{xy}$ strain---are all related by a single SVCR:
\begin{equation}
    (\partial_x^2 +\partial_y^2) \varepsilon_{A_{1g}}(\boldsymbol{r}) = (\partial_x^2 -\partial_y^2) \varepsilon_{B_{1g}}(\boldsymbol{r}) + (2\partial_x\partial_y) \varepsilon_{B_{2g}}(\boldsymbol{r}), \label{eq:SVCR_2D}
\end{equation}
which is straightforwardly proven from the definition of the infinitesimal strain tensor. Whereas the dilatation is a symmetry-preserving, volume-changing strain, both the deviatoric and shear are symmetry-breaking, but volume-preserving, strains. The SVCR above shows that mesoscopic shear ($\varepsilon_{B_{2g}}$) modulations incur an additional elastic energy associated with dilatation ($\varepsilon_{A_{1g}}$)  and thus related to the bulk modulus, unless they are waves with momentum lying along the $[100]_T$  or $[010]_T$ axes. We demonstrate this constraint in Fig.~\ref{fig:shear_waves}, where we show that the interdependence between the shear and dilatation waves is controlled by the momentum direction. In Fig.~\ref{fig:shear_waves}\textbf{(a,c)}, a compatible shear wave with momentum along $[100]_T$ induces a displacement vector that has no dilatation strain, whereas when the momentum is along $[110]_T$, the associated displacement vector triggers local volume changes that maximize the dilatation strain (see Supplementary Information).

Whereas the strain basis is adopted from the crystalline point group $\text{D}_{4h}$,  Eq. \eqref{eq:SVCR_2D} is a fundamental constraint imposed by geometry, and therefore applies for media of any symmetry or elastic stiffness \citep{Meese_short_paper_2025, Meese_long_paper_2025}. Thus, there is a particular universality for strain waves in materials near tetragonal-to-orthorhombic ferroelastic transitions. Regardless of the specific elastic properties of a tetragonal crystal, $\varepsilon_{xy}$ spatial fluctuations above the ferroelastic transition will generally induce energetically expensive dilatations unless the wave vector lies along the in-plane crystal axes. The softest fluctuations above the transition will then also be those without dilatations and the modes with momenta pointing along the in-plane crystal axes will be the first to freeze into static modulations below the transition \citep{paulLatticeEffectsNematic2017, Meese_short_paper_2025, Meese_long_paper_2025}.

\section*{\texorpdfstring{\hspace*{-\parindent}Proposed mechanism for the strain waves in Cu-Ba122}{Proposed mechanism for the strain waves in Cu-Ba122}}\label{sec:theory}

The SVCR in Eq. \eqref{eq:SVCR_2D} provides a general geometric reason for why we would expect $B_{2g}$ strain waves to modulate along the $[100]_T$/$[010]_T$ axes if such modulations do exist in the crystal. Here, we further propose a minimal real-space elastostatic model that attempts to address why it would be energetically favorable for periodic modulations to arise in the examined twin domains in the first place. Without any pretense that this model indeed is the exact explanation for our observations, we show that if the twin is modeled as a one-dimensional object, restricted in direction by the SVCR, then uniform twin domains with a sharp twin boundary are unstable toward spatial modulations. Moreover, the characteristic modulation wavelength $\Lambda$ emerges as a length scale almost an order of magnitude larger than the mean-field correlation length of the order parameter, $\xi_{\mathrm{MF}}$, yet is still orders of magnitude smaller than the size of the twin, $L$, away from the critical region.

First and foremost, the restriction of the modulation momentum to the in-plane crystal axes renders the two-dimensional nemato-elastic problem effectively one-dimensional. For concreteness, we consider modulations along the $[100]_T$ direction. For modulations along this direction, there is a local proportionality between the shear strain, $\varepsilon_{xy}(\boldsymbol{r})$, and the $B_{2g}$ electronic nematic order parameter, $\phi(\boldsymbol{r})$. Consequently, the complexity associated with solving coupled nonlocal equations for spatially modulated strain and nematic fields is reduced to only solving for a single field in the 1D limit (see Supplementary Information).

In the absence of external strain fields, the traction-free boundary conditions from elasticity theory (see Supplementary Information) coupled with nemato-elasticity require that $\phi(x) \rightarrow 0$ on the outer boundaries of the twins, modeled here at $x = 0$ and $x = L$. The twinning, however, adds the constraints that the electronic nematic order parameter is finite and of opposite signs within the bulk of each twin component and goes through zero at the twin boundary. Thus, we are left to solve a constrained Ginzburg-Landau minimization problem along the $x$-axis, modeling the nematic free energy as the usual $\phi^4$ model with the following 1D free energy functional:

\begin{equation}
    \mathcal{F}[\phi(x)] = \frac{1}{2}\int_0^L \text{d}x\, \left\{ a(T-T^*)\phi^2(x) + [\partial_x\phi(x)]^2 \right\} + u\int_0^L \text{d}x\, \phi^4(x),
\end{equation}
for field configurations, $\phi(x)$, satisfying the stated constraints. In the above expression, $L$ is the size of the two twin domains, $a >0$ is the coefficient of the inverse susceptibility as a function of temperature, $T$, $T^*$ is the nematic critical temperature, and $u>0$ provides thermodynamic stability. The free energy is given in units where the nematic stiffness is one. Given that the stated boundary conditions over-determine the problem, we use a variational approach to obtain a metastable \textit{ansatz} that admits modulations within the twin domains.

From a macroscopic perspective, we expect twin domains to be uniform within the bulk, separated by sharp domain walls. Intuitively, a square wave solution satisfies the prescribed boundary conditions and should be thermodynamically stable. We therefore choose the physically plausible \textit{ansatz} of a partial Fourier series of the square wave, which both has the expected limiting behavior and allows for spatial modulations:
\begin{equation}
    \phi_M(s) \equiv \Phi \cdot \frac{4}{\pi}\sum_{m \; \text{odd}}^M \frac{\sin(2\pi m s)}{m},\;s\equiv \frac{x}{L}\in (0,1),
\end{equation}
where the amplitude, $\Phi$, and the upper bound of the summation, $M$, are variational parameters. The first limit, $M = 1$, corresponds to a single sinusoid of wavelength $L$. The second limit, $M\rightarrow\infty$, meanwhile, corresponds to flat twin domains with an infinitely sharp domain wall between the two. Whereas neither limit has spatial modulations within either twin domain, any $M \in (1, \infty)$ solution will contain bulk modulations. As shown in the Supplementary Information, the value of $M$ that minimizes the free-energy functional is finite below the transition temperature, indicating that spatially modulated twin domains are energetically favored over uniform twin domains separated by a sharp boundary. Moreover, we show that our model predicts both the wavelength and amplitude of the spatial modulations to slowly decrease within the ordered phase, scaling as $1/\sqrt{T^* - T}$.
Since our data were collected at temperatures not in the immediate vicinity of $T_S \simeq 94$ K, the model is consistent with our observation that the modulation wavelength does not change appreciably as a function of the probed temperatures.

\section*{\texorpdfstring{\hspace*{-\parindent}Discussion}{Discussion}}\label{sec:discussion}

Our results demonstrate the power of DFXM as a novel tool to probe mesoscopic strain fields in quantum materials, in which structural distortions can have important effects on functional, electronic, and magnetic properties. Mesoscopic phenomena are prevalent in strongly-correlated systems~\cite{Dagotto2005ComplexitySystems}, with strain and phase separation affecting transport~\cite{Lai2010MesoscopicFilm} and superconducting behavior~\cite{Jin2024First-PrinciplesStructure} of a variety of complex oxides. With the installation of fourth-generation synchrotrons, improvements in X-ray optics, and development of novel depth-sensitive imaging techniques such as coded-aperture imaging~\cite{Gursoy2025Dark-fieldImaging}, DFXM will be well-positioned to shine new light on the structural aspects of many complex materials.

Our measurement also opens new real-space perspectives on the relation between the electronic and lattice degrees of freedom in iron-based superconductors. These materials exhibit strong nemato-elastic coupling~\cite{Lahiri2022Defect-inducedMaterials}, and the pronounced elastoresistive response reported in prior measurements~\cite{Kuo2013MeasurementBaFe0.975Co0.0252As2, Kuo2016UbiquitousSuperconductors} suggests that the strain wave amplitude of $10^{-5}$ observed in our results could induce significant spatial variations in the local resistivity anisotropy within individual domains. However, it remains unknown whether, or how, this phenomenon influences superconductivity in these materials. As the orthorhombic order parameter is suppressed by increased doping, our model predicts a corresponding suppression of the strain wave amplitude within domains. Intriguingly, in Ba(Fe$_{1-x}$Co$_x$)$_2$As$_2$, it has been shown that as doping approaches the optimal value for superconductivity from the underdoped regime, the nemato-elastic coupling weakens while the nematic scattering cross section increases dramatically~\cite{Ikeda2021ElastocaloricFluctuations}. Given that superconducting pairing is strongly enhanced by nematic fluctuations near a nematic quantum critical point (QCP)~\cite{Lederer2015EnhancementPoint}, and that strain can suppress superconductivity by damping these fluctuations~\cite{Malinowski2020SuppressionPoint}, investigating how such strain waves evolve with doping is essential for understanding their impact on superconductivity. 

\newpage

\bibliography{references, references-2}

\section*{Methods}\label{sec:methods}

\subsection*{Growth and characterization of single crystals}

Single crystals of Ba(Fe$_{1-x}$Cu$_x)_2$As$_2$ were grown using an FeAs flux as described elsewhere \cite{Chu2009DeterminationBaFe1-xCox2As2, Ni2010TemperatureCrystals}. The crystals grow with a plate-like morphology, where the large faces correspond to the $ab$~plane of the tetragonal lattice. The sample used for the DFXM measurements in this work was taken from a growth batch with a Cu composition of $x = 0.02 \pm 0.0008$ as determined by electron probe microanalysis wavelength-dispersive
spectroscopy (EPMA-WDS) using parent compound BaFe$_2$As$_2$ and elemental Cu metal as calibration samples. The measured sample was cleaved from a single crystal using a razor blade, and it had the dimensions 3.9 mm $\times$ 1.4 mm $\times$ 85 $\mu$m.

We established the mean and standard deviation of the Cu concentration of the batch by performing WDS on eleven freshly-cleaved samples chosen from various single crystals within the batch in a JEOL JXA-8230 “SuperProbe” electron microprobe. Ten distinct $\sim$1~$\mu$m-sized spots across each of the eleven samples were selected for measurements to estimate the spatial and statistical variation of Cu concentration within individual samples and across the batch. No systematic spatial variation of Cu concentration was detected and the standard deviation was found to be $0.0008$ across eleven distinct samples with a mean Cu concentration of $x = 0.02$.

Four-point resistivity measurements were performed on five samples from the same batch to estimate the mean transition temperatures $T_S$ and $T_N$ for the structural transition and the Neel transition of the batch, respectively. It has been experimentally shown that the Fisher-Langer relation $C_p^{(c)} \propto \frac{\partial \rho^{(c)}}{\partial T}$ between the critical anomaly in the heat capacity and that in the temperature derivative of resistivity holds in the vicinity of both second-order phase transitions in this material class \cite{Hristov2019ElastoresistivePoints}. From the resistivity derivative measurements, we determined that for this batch $T_S = 94.1\,\pm\,1.5$ K and $T_N = 85.7\,\pm\,1.7$ K. The sample investigated in this paper was also confirmed to have $T_S = 94\,\pm\,1.5$ K using XRD as elaborated below.

\subsection*{High-energy X-ray diffraction (XRD)}

Prior to measuring the sample in DFXM, we performed high-energy X-ray diffraction on the sample at Beamline 6-ID-D of the Advanced Photon Source, Argonne National Laboratory, USA. The sample was glued from its corner on a copper mount with its $[100]$ crystal axis aligned vertically and positioned on a Huber six-circle diffractometer such that a large free-standing sample volume was at the center of rotation. We illuminated the sample with a monochromatic X-ray beam with beam size $500 \times 500$ $\mu$m$^2$ and X-ray energy of 87 keV. A Dectris Pilatus 2M CdTe area detector was positioned 2.6 m away from the sample and off the beam axis to capture the tetragonal 220 peak with high resolution. The sample was kept under vacuum inside a Kapton dome and cooled from 300 K down to 10 K using a cryocooler in steps of 2 K. At each temperature step, the sample was rotated around the vertical axis and the diffracted light was collected. Raw detector images were used to establish that the $220$ peak splitting occured at $T_S \simeq 94$ K.

\subsection*{Dark-field X-ray microscopy (DFXM)}

We performed the DFXM measurements at Beamline 6-ID-C of the Advanced Photon Source. The sample was mounted on the sample holder of a customized low-vibration Montana Instruments s100 helium cryostat, which itself was mounted on a diffractometer via a vibration-damped optical breadboard \cite{Plumb2023DarkNaMnO2}. The sample was kept under vacuum in a chamber with beryllium entry and exit windows aligned for a horizontal scattering geometry. The sample was glued onto the sample holder by GE varnish from one of its corners such that an area of $\sim$3.2 mm~$\times$~$1$ mm was free standing and available for measurement in the transmission geometry. We performed measurements on a volume of this free-standing part of the sample $\sim$$2.5-2.7$~mm away from the glued part, therefore, far away from any region that might be affected by the differential thermal contraction from the sample holder. The sample was aligned such that the $[110]$ direction of the crystal axes lay within the horizontal scattering plane.

We used a monochromatic beam of X-rays tuned to 20 keV by a Si (111) double-crystal monochromator to illuminate our sample. At this energy, the X-ray attenuation length estimated for BaFe$_2$As$_2$ is $\sim$80~$\mu$m, which is approximately the thickness of our sample. The X-ray beam was focused on the sample using a Beryllium compound refractive lens (CRL) resulting in an illuminated area of $\sim$$35 \times 35$~$\mu$m$^2$ on the sample with high photon flux density. Downstream of the sample a 50-$\mu$m pinhole and a polymeric CRL were positioned on the $2\theta$ arm of the diffractometer to image the light diffracted at a particular scattering angle while blocking out other divergent scattered light. The polymeric CRL was designed for use at 20 keV and has a focal length of 131 mm at that X-ray energy \cite{Qiao2020AOptics}; it acted as our objective lens and had a working distance of 140 mm from the sample. Another 50-$\mu$m pinhole was positioned at the back focal plane (BFP) of the objective lens to filter out light originating from off-angle scattering. The small numerical aperture of the objective lens in combination with the pinholes led to an angular resolution of $\sim$$0.001^\circ$.

We used a high-resolution area detector placed $\simeq$2.435 m downstream from the sample at the image plane to collect the X-ray microscopy images. The detector is composed of a 10~$\mu$m thick LuAG:Ce scintillator that converts X-rays to visible light, a 5$\times$ optical objective lens, and an Andor 5.5-megapixel sCMOS Zyla camera. At this configuration, the X-ray magnification from the poymeric CRL objective lens was 26$\times$ which, combined with the detector's 5$\times$ optical magnication, led to a total magnification of 130$\times$. Given the $6.5\times6.5$ $\mu$m$^2$ pixel size of the Zyla detector, this total magnification enabled us to have an effective pixel size of 50 nm for our microscopy images. The resolution of the microscope was determined to be 300 nm.
The field of view was determined by the illuminated area and had a size of $\sim$$35 \times 35$ $\mu$m$^2$.

\section*{Acknowledgments}
The authors would like to thank A. Kapitulnik and S. A. Kivelson for fruitful discussions. The work at Stanford University (crystal synthesis and characterization, data analysis) by K.A.Y., A.G.S., and I.R.F. received support from the DOE, Office of Science, BES, under contract DE-AC02-76SF00515. The authors would like to thank the Karlsruhe Nano Micro Facility (KNMF)  for the fabrication of the polymer X-ray optics. This research used resources of the Advanced Photon Source, a U.S. Department of Energy (DOE) Office of Science user facility operated for the DOE Office of Science by Argonne National Laboratory under Contract No. DE-AC02-06CH11357.

\section*{Author contributions}
K.A.Y. synthesized and characterized single crystals. K.A.Y., E.K., and Z.I. performed the DFXM measurements. K.A.Y., A.G.S., and M.J.K. performed the XRD measurements. K.A.Y. analyzed the raw DFXM and XRD data with assistance from E.K., M.J.K, and Z.I. K.A.Y., E.K., Z.I., and I.R.F. interpreted the experimental results. W.J.M. and R.M.F. developed the theoretical model. K.A.Y., W.J.M., and I.R.F. wrote the manuscript with contributions and discussions from all authors. Z.I. and I.R.F. supervised the project.

\clearpage

\begin{figure}[p]
    \centering
    \includegraphics[width=\linewidth]{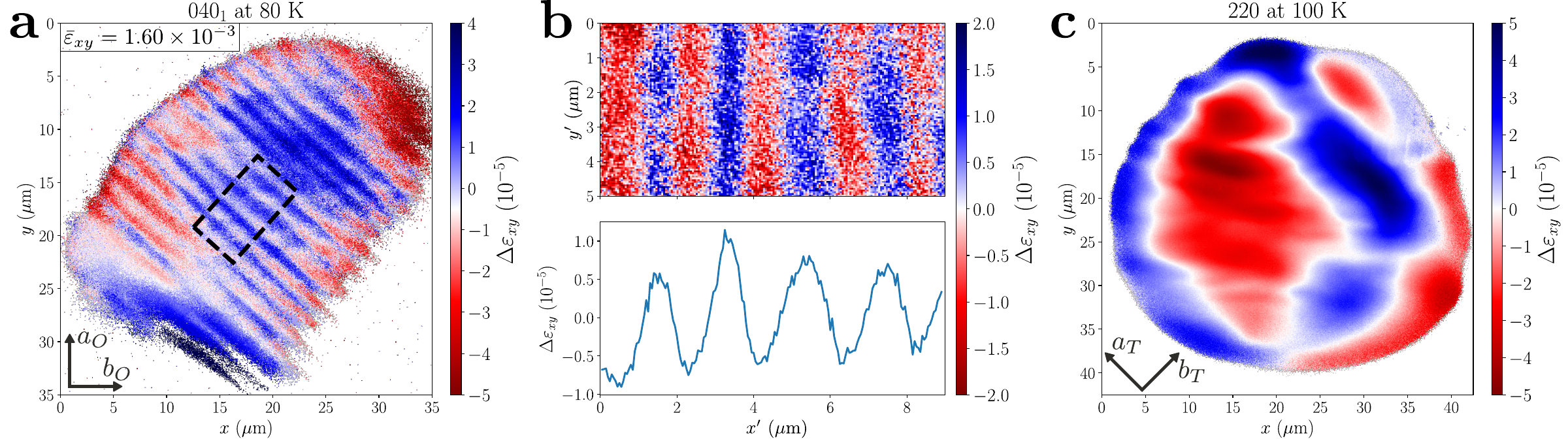}
    \caption{\textbf{Detection of strain wave within a single nematic domain in the orthorhombic phase and loss of periodicity in the tetragonal phase}
    \newline
    \textbf{a} Real-space strain map of a nematic domain scattering at the $040_1$ peak at 80 K. $\Delta\varepsilon_{xy}$ refers to the relative strain with respect to the median $d_{040}$-spacing value of the depicted volume. $\bar{\varepsilon}_{xy}$ is the average orthorhombicity measured at 80~K using conventional XRD. ${xy}$ subscript corresponds to the diagonal direction of the $ab$ face of the high-temperature tetragonal crystal structure. Dashed lines enclose a central area unaffected by vignetting effects at the boundaries and is shown in more detail in panel b. 
    \textbf{b} Magnified view of the area enclosed by the dashed lines in panel a. The primed axes $(x',y')$ are rotated by 45$^\circ$ with respect to the unprimed axes $(x,y)$.  $\Delta\varepsilon_{xy}$ is with respect to the median $d_{040}$-spacing value of the enclosed area. Plot below shows the values of $\Delta\varepsilon_{xy}$ averaged along the $y$ direction of the above plot. We observe that the ${xy}$ strain oscillates by $\sim$$10^{-5}$ with a wavelength of $2$-$3$~$\mu$m. 
    \textbf{c} Strain map of the $220$ peak in the tetragonal phase at $T = 100$~K. The spatial variation of strain lacks the coherence and periodicity observed in the nematic phase. 
    }
    \label{fig:fig3}
\end{figure}

\begin{figure}[p]
    \centering
    \includegraphics[width=\linewidth]{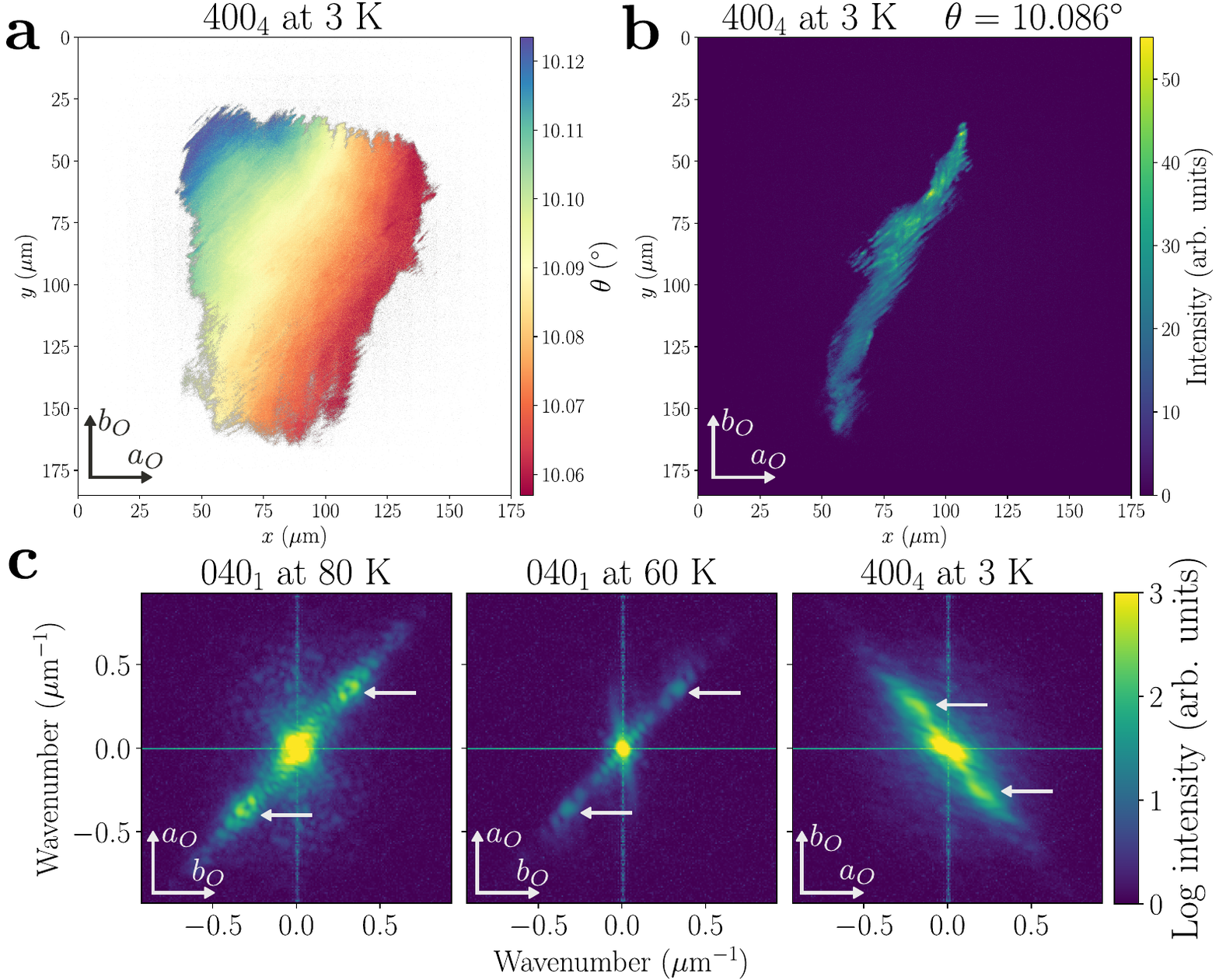}
    \caption{\textbf{Lack of lattice tilt contribution to intensity modulations and negligible dependence of wavelength on temperature}
    \newline
    \textbf{a} Large area spatial map of local orientation of the $(400)_4$ lattice planes as determined by the center of mass of the intensity distribution as a function of $\theta$ at each individual detector pixel. $\theta$~varies by $\sim 0.06^\circ$ over the large length scale of $\sim 100$~$\mu\textrm{m}$ and there are no observable orientation modulations with wavelength 2-3~$\mu\textrm{m}$.  
    \textbf{b} Large area map of scattering intensity at a particular lattice orientation value $\theta=10.086^\circ$. Intensity modulations with wavelength 2-3~$\mu\textrm{m}$ are visible across a length scale of $100$~$\mu\textrm{m}$. This intensity map is part of the dataset used to determine the local orientation map of panel \textbf{a}, and demonstrates that intensity modulations are observable in a region where orientation does not change appreciably.
    \textbf{c} Fourier transforms of microscope images of the $040_1$ and $400_4$ peaks at various temperatures. The white arrows highlight the main satellite peaks corresponding to the wavenumber of the intensity modulations.
    }
    \label{fig:fig4}
\end{figure}

\begin{figure}[p]
    \centering
    \includegraphics[width=0.645\linewidth]{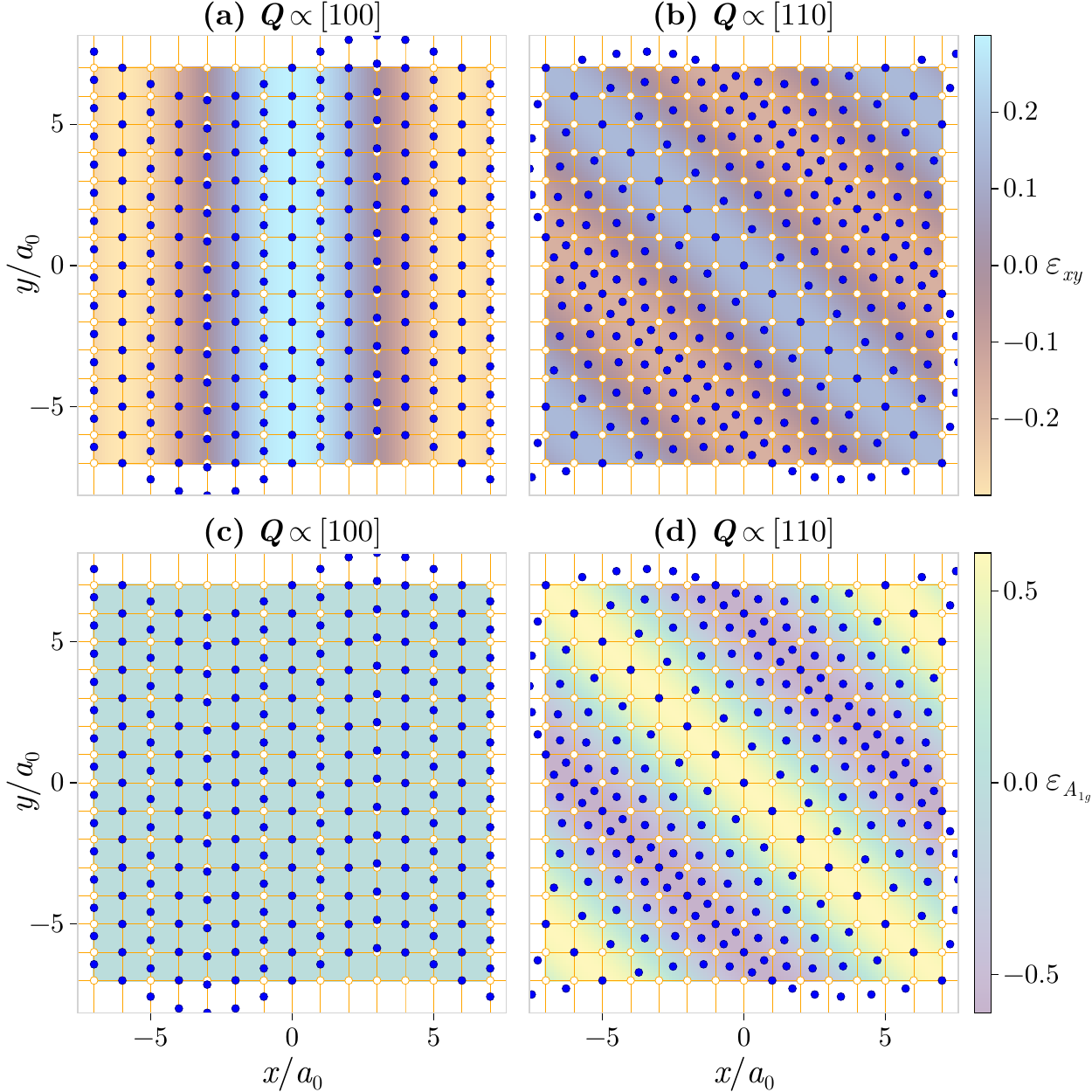}
    \caption{\textbf{Static shear waves constrained by Saint Venant Compatibility Relations.} 
    \newline
    A square lattice, with equilibrium positions (orange, open circles) and lattice constant $a_0$, which is deformed by a static shear wave $(\varepsilon_{xy})$ into displaced positions (blue closed circles). By satisfying the Saint Venant Compatibility Relation in Eq.~\eqref{eq:SVCR_2D}, the shear waves with arbitrary momenta will result in lattice displacements that may simultaneously induce volume-changing, energetically costly, dilatation strain: $\varepsilon_{A_{1g}} \equiv \varepsilon_{xx} + \varepsilon_{yy}$. \textbf{(a,b)} Shear waves with wave vector~$\boldsymbol{Q}\propto[100]$, and~$\boldsymbol{Q}\propto[110]$, respectively, and \textbf{(c,d)} the corresponding dilatation waves generated by them. The pair \textbf{{(a,c})} (also the pair \textbf{(b,d)}) has exactly the same displacement vector field, $\boldsymbol{u}(\boldsymbol{r})$, integrated from a compatible shear wave, $\varepsilon_{xy}(\boldsymbol{r})\propto \cos(\boldsymbol{Q}\cdot\boldsymbol{r})$. When $\boldsymbol{Q} \propto [100]$ \textbf{(a,c)} the dilatation strain vanishes, whereas when $\boldsymbol{Q} \propto [110]$ \textbf{(b,d)} the dilatation is maximal. The strain waves in this figure are depicted with a zero uniform shear strain background, whereas the nematic domains in the material studied have uniform shear strain values~($\sim$$10^{-3}$) much larger than the amplitude of the measured modulation~($\sim$$10^{-5}$). As discussed in the main text, only the non-uniform part of the shear strain induces a dilatation. The distortions depicted in this figure are exaggerated with respect to the lattice constant for clarity ($\varepsilon_{xy}\sim 0.2$ as opposed to the measured $\varepsilon_{xy}\sim 10^{-5}$ in Fig.~\ref{fig:fig3}). 
    }
    \label{fig:shear_waves}
\end{figure}

\clearpage

\makeatletter
\renewcommand{\theequation}{S\arabic{equation}}
\setcounter{equation}{0}
\renewcommand{\thefigure}{S\arabic{figure}}
\setcounter{figure}{0}
\renewcommand{\thepage}{S-\arabic{page}}
\setcounter{page}{1}
\renewcommand{\thesection}{S-\Roman{section}}
\setcounter{section}{0}
\setcounter{subsection}{0}

\begin{center}
    \textbf{\large Supplementary information: Discovery of spontaneous mesoscopic strain waves in nematic domains using
dark-field X-ray microscopy}
\end{center}

\begin{center}
    Kaan Alp Yay,$^*$
    W. Joe Meese,
    Elliot Kisiel,
    Matthew J. Krogstad,
    Anisha G. Singh,
    Rafael M. Fernandes,
    Zahir Islam,$^\dagger$
    Ian R. Fisher$^\ddag$
\end{center}

\begin{center}
    Corresponding authors: $^*$kaanalpyay@stanford.edu, $^\dagger$zahir@anl.gov,
    $^\ddag$irfisher@stanford.edu
\end{center}

\clearpage

\section{Reconstruction of DFXM data}\label{sec:Supp1_data_reconstruction}
\subsection{\label{subsec:pre-processing}Pre-processing and noise removal}

All raw images from the Zyla camera in .tiff file format were pre-processed using (1) background subtraction followed by (2) cropping to region of interest, (3) threshold removal, (4) removal of isolated pixels. Most of the $2558 \times 2158$ pixels of the camera for a given image did not contain a diffraction signal, therefore we used the median value of the intensity distribution of all pixels as the background value for a given image. This background value remained static between different sets of images and was subtracted from each image. The images were then cropped to an empirically determined region of interest to reduce the computational cost of further analysis. In the cropped image all pixel values below an empirically determined threshold were set to zero to remove random noise. Finally, isolated pixels with high intensity whose neighboring pixels all have zero intensity were also set to zero. Our pre-processing methodology follows similar steps to the \textit{darfix} package in Python~\cite{GarrigaFerrer2023DarfixMicroscopy}.

\subsection{\label{subsec:blurriness}Selection of images with sharpest features from a particular scan}

\begin{figure}[ht]
    \centering
    \includegraphics[width=\linewidth]{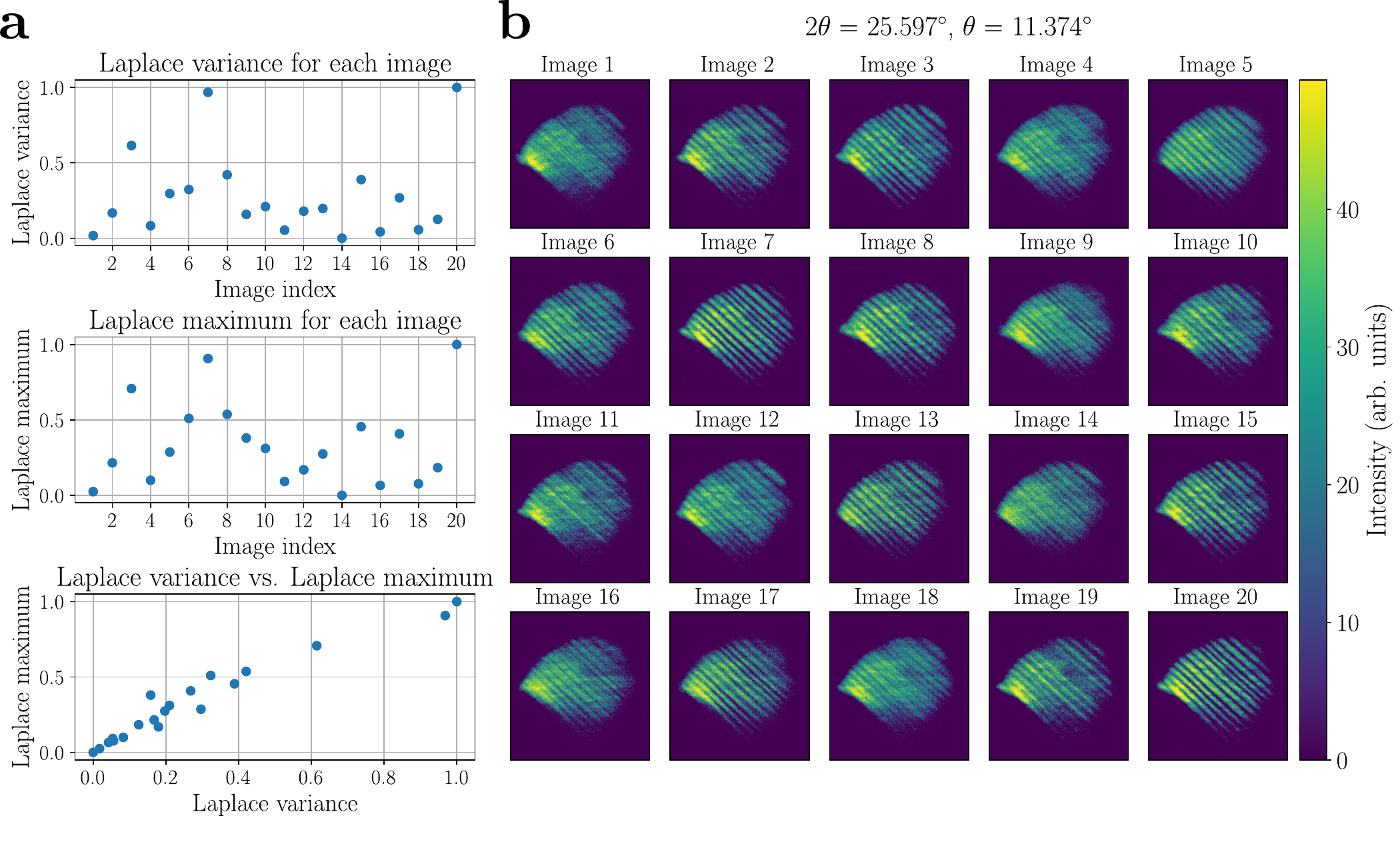}
    \caption{\textbf{Automatic detection of the sharpest images from a dataset}
    \newline
    \textbf{a} Laplacian variance and Laplacian maximum distribution for a dataset of 20 images all collected consecutively at the same $(\theta,2\theta)$ value. The lowest panel shows that these two metrics for quantifying image sharpness track each other almost linearly.
    \textbf{b} The images for which the sharpness metrics are plotted in panel \textbf{a}. We can confirm that images determined to be sharp by the Laplacian metrics are indeed visually the sharpest.
    }
    \label{fig:automatic_detection}
\end{figure}

Our datasets consisted of (1) $\theta$ scans where we stepped over individual $\theta$ motor values within a designated angular interval at a fixed $2\theta$ value, or (2) $\theta$-$2\theta$ scans where we iterated over a nested loop of $\theta$ and $2\theta$ motor values. For each $(\theta, 2\theta)$ value in either of these two types of scans, we took multiple images for redundancy. An example dataset of images taken consecutively at a fixed $(\theta, 2\theta)$ value within the $040_1$ peak at 80~K can be seen in Fig.~\ref{fig:automatic_detection}\textbf{b}. As can be seen in this example dataset, some images within the set have sharper features whereas the others have more blur, which we ascribe to the small yet present vibrations coming from the compressor of the cryostat. 

To automatically select the sharpest image from a given dataset, we used a method involving the Laplacian operator of the Open Source Computer Vision (OpenCV) library. The Laplacian operator approximates the second derivative of the pixel intensity with respect to pixel position and is therefore sensitive to detecting sharp edges. Therefore, the variance and the maximum value of the Laplacian of a given image are commonly used metrics in computer vision to quantify the sharpness of that image: the higher the variance or the maximum of the Laplacian, the sharper the image \cite{Bansal2017BlurOpen-CV, Pertuz2013AnalysisShape-from-focus}.

The algorithm we used to detect the sharpest image from a dataset is as follows:
\begin{enumerate}
    \item Smoothen each image by convolving it with a Gaussian filter to reduce noise, (using the GaussianBlur function of OpenCV)
    \item Apply Laplacian operator to each image and compute the variance and maximum value of the resulting Laplacian images,
    \item Normalize the Laplacian variance and maximum values for each image with respect to the difference between their maximum and minimum values within a given dataset. So, for Var$L(i)$ being the Laplacian variance and Max$L(i)$ being the maximum value of the Laplacian of image $i$:
    $$
    \mathrm{Var}L(i)_{\text{norm}}
    = \frac{\mathrm{Var}L(i) - \text{min}_i(\mathrm{Var}L(i))}
    {\text{max}_i(\mathrm{Var}L(i)) - \text{min}_i(\mathrm{Var}L(i))}
    $$
    $$
    \mathrm{Max}L(i)_{\text{norm}}
    = \frac{\mathrm{Max}L(i) - \text{min}_i(\mathrm{Max}L(i))}
    {\text{max}_i(\mathrm{Max}L(i)) - \text{min}_i(\mathrm{Max}L(i))}
    $$
    \item Select the sharpest image by determining the image with the highest quadrature sum of normalized Laplacian variance and maximum values:
    $$
    i_{\mathrm{sharpest}} = \mathrm{arg max}_i \left\{(\mathrm{Var}L(i)_{\text{norm}})^2 + 
    (\mathrm{Max}L(i)_{\text{norm}})^2\right\}
    $$
\end{enumerate}

As can be seen in the lowest panel of Fig.~\ref{fig:automatic_detection}\textbf{a} the normalized Laplacian variance and Laplacian maximum values for images from the example dataset correspond to each other in an approximately linear fashion. Our algorithm amounts to selecting the image with the largest radius in this two-dimensional space of sharpness metrics. A visual comparison of the sharpness metrics of each image in Fig.~\ref{fig:automatic_detection}\textbf{a} with the actual images in Fig.~\ref{fig:automatic_detection}\textbf{b} confirms that the used metrics indeed distinguish the sharpest images (20 and 7) from blurrier ones. In this dataset, the algorithm would select image 20 as the sharpest image for further analysis. We empirically confirmed the success of this algorithm in selecting the sharpest images in several other datasets.

\subsection{\label{subsec:strain_scan}Image registration and construction of strain maps from \texorpdfstring{$\theta$--$2\theta$}{theta--2theta} scans}

To construct the strain map, we selected images whose $(\theta,2\theta)$ values lie on a line cut  along the axis of the scattering vector $\vec{Q}$ through the investigated Bragg peak. The intensity in these images correspond to diffraction coming from parts of the illuminated area of the sample with different $d$-spacing (strain) values.

Because our detector is stationary during a $\theta$--$2\theta$ scan, a change in $2\theta$ leads to a finite shift of images along both the horizontal and vertical directions of the detector pixel array. Therefore, image registration is necessary to assign pixels from different images to the correct locations on the sample. Our image registration process consisted of (1) determining the $(x,y)$ shift between successive images by calculating their cross-correlation using the 2D discrete Fourier transform (fft2) function of NumPy, and (2) shifting the images according to their determined shift value using the ndimage.shift function of SciPy, which performs shifts with subpixel accuracy using a cubic spline interpolation. We empirically confirmed that the $(x,y)$ shift between different images are linear as a function of $2\theta$.  

After the shift correction and image registration, we assigned each pixel $(x,y)$ its corresponding $2\theta$ value by determining the center of mass of the intensity distribution as a function of $2\theta$. The corresponding $d$-spacing value for each pixel was then obtained using Bragg's law, and the relative strain $\Delta\varepsilon_{xy}$ was calculated by finding the relative difference in the $d$-spacing at each pixel from the median $d$-spacing value in the imaged region.
\begin{equation}
    d = \frac{\lambda}{2\sin \frac{2\theta}{2}}
\end{equation}
\begin{equation}
    \Delta\varepsilon_{xy} = \frac{d - d_{\mathrm{med}}}{d + d_\mathrm{med}}
    \approx \frac{d - d_{\mathrm{med}}}{2 d_\mathrm{med}}
\end{equation}
We used the notation $\Delta\varepsilon_{xy} = \Delta\varepsilon_{B_{2g}}$ to describe the variation of the $d$-spacing of both the $(220)$ lattice planes in the high-temperature tetragonal phase (Fig.~\ref{fig:fig3}\textbf{c}) and the $(040)_1$ lattice planes in the orthorhombic phase (Fig.~\ref{fig:fig3}\textbf{a}) to emphasize the $B_{2g}$ character of the spontaneously developing electronic nematicity in this material, which is defined with respect to the $D_{4h}$ point group of the parent tetragonal phase.

\subsection{\label{subsec:mosaicity_scan}Construction of large area local orientation maps from \texorpdfstring{$\theta$}{theta} scans}

We constructed the large area maps shown in Fig.~\ref{fig:fig3} by performing a fine $\theta$ scan at 100 different locations in the sample (a $10\times 10$ array of $(x,y)$ positional values on the sample). To determine the variation of the diffraction intensity for a given $\theta$ value across the whole $\sim 100$ $\mu$m $\times$ $100$ $\mu$m area, we shifted each image in the array by the known shift value of the motor moving the sample. The pixel intensities of adjacent images were averaged for the regions in which they overlapped. An example large area diffraction intensity map for $\theta = 10.086^\circ$ is given in Fig.~\ref{fig:fig3}\textbf{b}. The local orientation map for the large area was then calculated by finding the center of mass of the intensity distribution at each individual pixel as a function of $\theta$, and the resulting $\theta$ value was assigned to that pixel. The resulting local orientation map is shown in Fig.~\ref{fig:fig3}\textbf{a}.

\newpage

\section{\label{sec:fourier}Fourier transform studies}

\subsection{\label{subsec:fourier_method}Generation of aggregate Fourier transform plots}

Each dataset used in the Fourier transform (FT) study consisted of 10,000--100,000 images taken at a given temperature. The datasets collected at 80~K and 60~K were part of a $\theta$--$2\theta$ scan of the $040_1$ peak with multiple images taken at each $(\theta,2\theta)$ value and the sample location investigated was fixed. The dataset collected at 3~K was part of a large area $\theta$ scan of the $400_4$ peak; a $\theta$ scan was performed at several different locations on the sample, and multiple images were taken for each $\theta$ value. To construct the Fourier transform plots displayed in Fig.~4\textbf{c}, we first selected a subset of the images taken at every $(\theta,2\theta)$ value in a given dataset and pre-processed them as described above. Then, we used the fft2 function of NumPy to determine the 2D FT of each image, and added the absolute value of the FTs of all images for every given $(\theta,2\theta)$ value. The absolute values of all the FTs of all $\theta$ values were then added for a particular $2\theta$ value at the center of the corresponding Bragg peak. Finally the logarithm of the aggregate FT arrays was computed to create the FT plots shown in Fig.~4\textbf{c}. The procedure used to generate the plots was applied to all the other $2\theta$ values for the 80~K and 60~K datasets, and for multiple different locations on the sample for the 3~K dataset. 

\subsection{\label{subsec:fourier_wavelength}Wavelength determination}

Once the aggregate FT plots were generated, we determined the average wavelength corresponding to each $2\theta$ value and sample location using an automated procedure. First, to quantify the angular orientation of satellite peaks in a two-dimensional FT, we developed a method that identifies and analyzes high-intensity features in the FT image. The FT was first filtered using a cross-shaped mask to block out the central peak and detector artifacts along the $x$ and $y$ axes, and to enhance the visibility of satellite peaks. A dynamic thresholding scheme, based on the image's intensity standard deviation, was applied to isolate prominent peaks, and connected component labeling was used to identify and retain significant clusters of high intensity. Once two satellite peaks with the highest intensity were isolated, their centers of mass were calculated and a linear fit was performed to determine their angular deviation from the horizontal axis. The method returned the rotation angle of the peaks. This automated approach ensured robustness against noise and spurious features by adaptively refining the threshold until exactly two dominant satellite peaks were identified.

\begin{figure}[t]
    \centering
    \includegraphics[width=\linewidth]{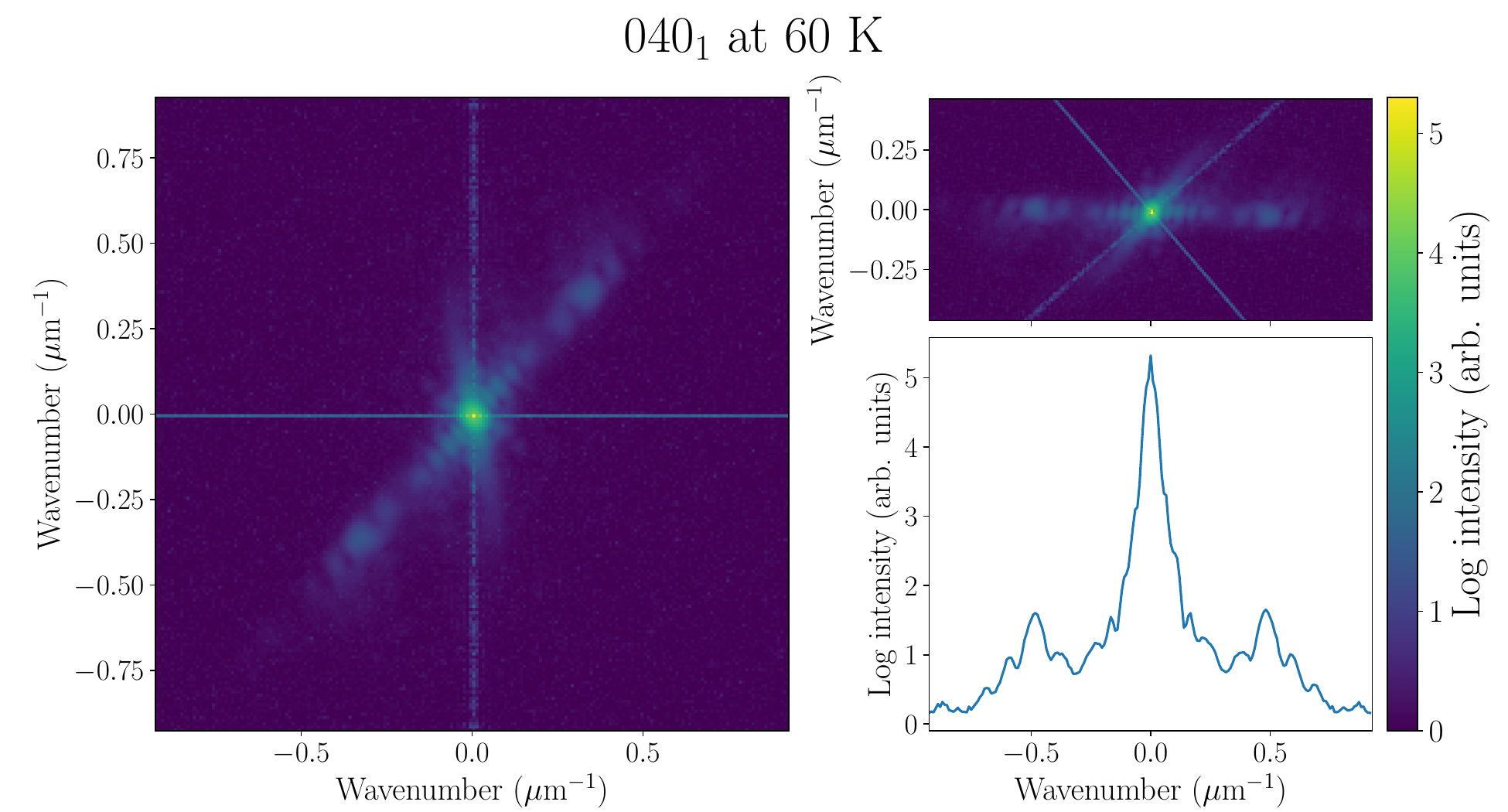}
    \caption{\textbf{Wavelength determination from Fourier transform
    plots}
    \newline
    The aggregate FT plot of the $040_1$ peak at 60~K for $2\theta = 25.611^\circ$ from Fig.~\ref{fig:fig4}\textbf{c} is recreated here. On the right panel, the FT plot rotated as described in the text is shown above and its average taken along the $y$-axis is shown below. The line plot clearly shows the main satellite peaks at $\sim$$0.5$~$\mu$m$^{-1}$ corresponding to a wavelength of $\sim$$2$~$\mu$m.
    }
    \label{fig:fourier_wavelength}
\end{figure}

The spectrum was then rotated according to the determined angle to align the peaks horizontally, and a region of interest was extracted and averaged along the $y$-axis to yield a one-dimensional intensity profile, as depicted on the right panel of Fig~\ref{fig:fourier_wavelength}.

To extract the real-space wavelength from the one-dimensional Fourier amplitude profile, satellite peaks were identified using the find\_peaks method of SciPy, a prominence-based peak-finding algorithm. The prominence threshold was adaptively tuned until exactly three peaks were detected, corresponding to the central and the two main satellite peaks expected from the periodic modulation. A Gaussian fit was then applied to a windowed region around the right satellite peak to determine its center with subpixel precision. The corresponding wavenumber was calculated in spatial frequency units ($\mu$m$^{-1}$), and the wavelength was obtained as its inverse in microns. If the satellite peaks could not be reliably detected due to falling out of the Bragg condition at certain $2\theta$ values or locations on the sample, the function returned zero.

\begin{figure}[t]
    \centering
    \includegraphics[width=\linewidth]{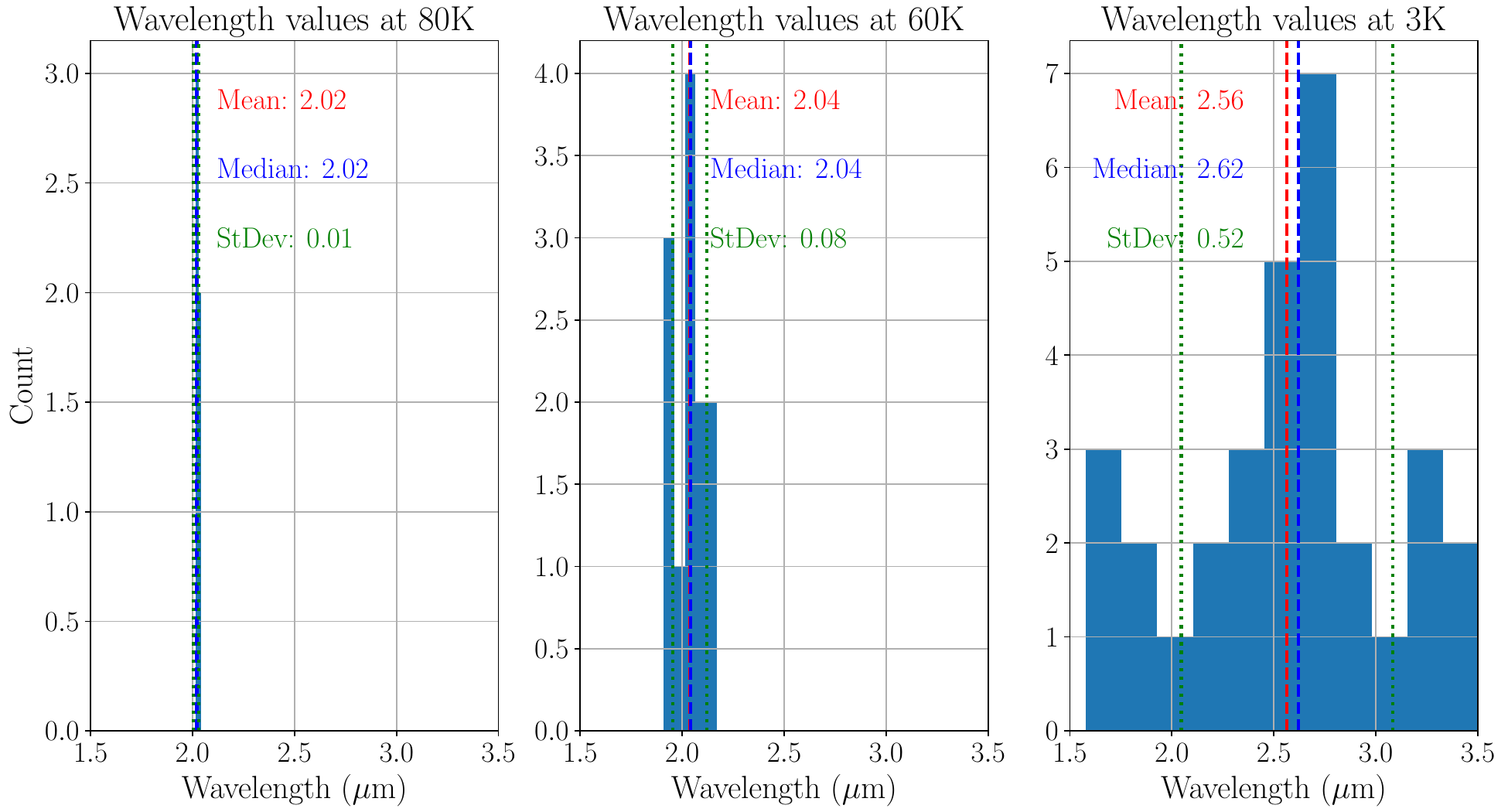}
    \caption{\textbf{Histograms of wavelength values determined at different temperatures}
    \newline   
    The count of wavelength values determined for a range of $2\theta$ values at 80~K and 60~K, and for range of sample locations at 3~K. The red (blue) dashed lines indicate the mean (median) value of each plot, and the green dotted lines indicate the standard deviation of the sampled data points. The limits of the $x$-axes for the plots are set to be the same to demonstrate the differing variance of wavelength values detected at each temperature. This difference is discussed in the text.
    }
    \label{fig:wavelength_histograms}
\end{figure}

The result of the described wavelength determination at differing $2\theta$ values for the datasets collected at 80~K and 60~K, and at differing sample locations for the dataset collected at 3~K are given in Fig.~\ref{fig:wavelength_histograms}. We collected the 80~K and 60~K data consecutively at the same location on the sample by performing $\theta$--$2\theta$ scans on the same peak $040_1$. We determined that the difference between the means of the wavelength values at the two temperatures was not significant given the variance in the datasets. The dataset at 3~K, however, has a mean wavelength value significantly larger than that of the other datasets. It is important to note that the 3~K dataset was collected by imaging the $400_4$ peak, and hence the observed modulations originated from a different domain than the one imaged at higher temperatures. The domain size itself potentially plays a role in the wavelength of the modulations. Thus, it is difficult to make a one-to-one comparison of the wavelength values measured in this domain versus the other to extract an empirical dependence of the wavelength of modulations on temperature. Nonetheless, the fact that the mean wavelength does not change significantly between 80~K and 60~K, and is of the same order of magnitude in the three datasets analyzed suggests that the wavelength of modulations varies weakly with temperature, and the two different domains investigated presumably are of roughly similar size.

In a related observation, the histograms for 80~K and 60~K show a small relative variation from the mean of the detected wavelength values as $2\theta$ is varied. In contrast, the histogram for 3~K shows a much larger relative variation from the mean value. We ascribe the larger variance in the 3~K dataset to two factors. First, we determined that the 3~K dataset consisted of blurrier images with inherently less sharp features; therefore, the FT analysis resulted in satellite peaks also with broader features, adding extra variance to the wavelength determination. In many images, the blurriness prevented our algorithm from detecting any significant peaks, and these images were not included in the wavelength analysis. Second, the 3~K dataset was collected for the $400_4$ peak at 100 different locations on the sample (a $10\times 10$ array of $(x,y)$ positions), whereas the 80~K and 60~K datasets were collected for the $040_1$ peak at a single location. As the 3~K dataset covers a larger area of the investigated domain, it potentially includes effects closer to domain walls or other external strain effects that can introduce added variance to the wavelength of the observed modulations.

\subsection{\label{subsec:subdominant}Observation of subdominant peaks in the Fourier transform plots}

\begin{figure}[t]
    \centering
    \includegraphics[width=0.8\linewidth]{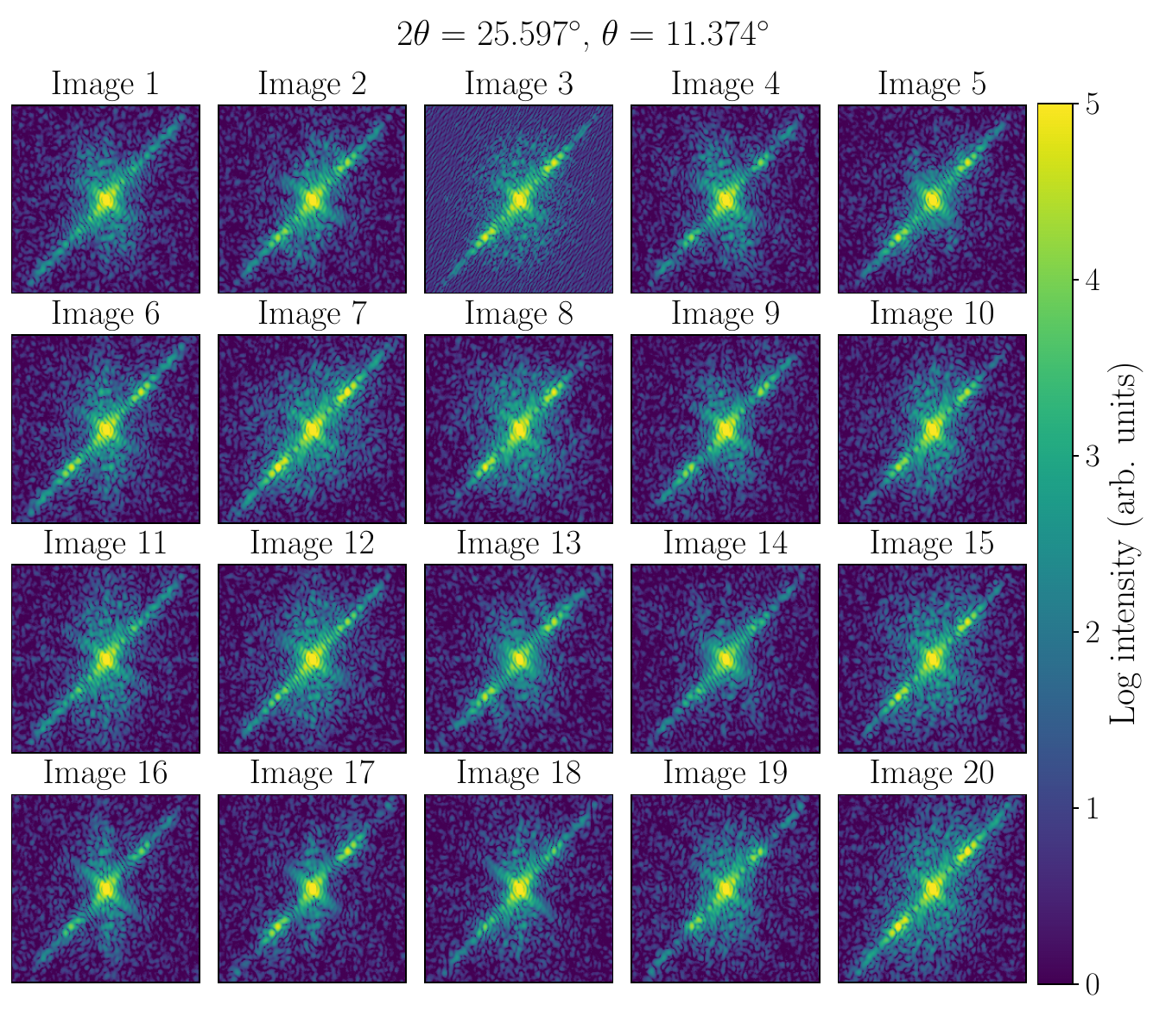}
    \caption{\textbf{Fourier transform of individual images from Fig.~\ref{fig:automatic_detection}\textbf{b}}
    \newline
    FT plots of individual images also show subdominant peaks along with the dominant satellite peaks and the central peak. The subdominant peaks are unidirectional along the dominant satellite peaks and are also periodic.
    }
    \label{fig:ft_individual_images}
\end{figure}

From Fig.~\ref{fig:fig4}\textbf{c} we notice that there are additional peaks in the FT plots with smaller intensity besides the dominant satellite peaks. These subdominant peaks are especially visible in the FT plots of the $040_1$ peak at 80 K and 60 K as these datasets consist of sharper images compared to the 3~K dataset as discussed above. As the FT plots at 60 K and 80 K in Fig.~\ref{fig:fig4}\textbf{c} consist of the sum of the FT of multiple images at several different $\theta$ values for a given $2\theta$ value, we also checked the FT plots of individual images to confirm that the subdominant peaks are inherent to all images and not merely an artifact of the aggregation procedure discussed above. The resulting FT plots of the individual images displayed in Fig.~\ref{fig:automatic_detection}\textbf{b} are shown in Fig.~\ref{fig:ft_individual_images}. As these FT plots demonstrate, the subdominant peaks are present for each image. Examining the FT plots of individual images in Fig.~\ref{fig:ft_individual_images} reveals that that the subdominant peaks are periodic. We ascribe this observation to a combination of two possible causes: (1) The pinhole in front of the objective lens acts as a circular mask on the image, and (2) there is a slight wavelength modulation approximately every five periods. While the cause of the wavelength modulation is currently unknown to us, both of these effects would lead to an FT spectrum with periodic sidebands around the main satellite peaks described by Bessel functions of the first kind, explaining the subdominant peak spectrum we observe.

\newpage

\section{Nemato-elasticity and shear strain waves}

In this section, we use a two-dimensional Ginzburg-Landau formalism to capture the interaction of the strain tensor and the electronic nematic order parameter in iron pnictide Ba(Fe$_{0.98}$Cu$_{0.02}$)$_2$As$_2$. We will show that the full nemato-elastic problem in two spatial dimensions with three interacting fields naturally leads to the effective one-dimensional problem described in the main text. We then show that a single sinusoid and a perfectly uniform square wave are both unstable towards the partial square-wave \textit{ansatz} taken in the main text. This holds within the nematic phase, with the solution vanishing completely above the nematic transition temperature. The minimizing partial wave solution for the twin has an accompanying temperature-dependent spatial modulation in the bulk of the twin components. Both the length scale and amplitude of the spatial modulation are slowly decreasing functions of temperature in the ordered phase, each scaling as $1/\sqrt{T^* -T}$.

\subsection{Development of spontaneous strain waves}

The parent compound, BaFe$_2$As$_2$, is well known to exhibit an Ising electronic nematic instability before the onset of stripe-antiferromagnetic order \citep{Chu2012DivergentSuperconductor, fernandesPreemptiveNematicOrder2012, fernandesWhatDrivesNematic2014, Fernandes2022IronSuperconductivity}. Meanwhile, the development of a nonzero, ferroelastic shear strain is a consequence of the symmetry-induced nemato-elastic bilinear coupling that the Ising nematic order parameter, $\phi$, has with the shear strain, $\varepsilon_{xy}$. The full nemato-elastic free energy then is the sum of three parts, and is written as
\begin{equation}
    F=F_{\text{n}}\left[\phi\left(\boldsymbol{r}\right)\right]+F_{\text{e}}\left[\varepsilon_{ij}\left(\boldsymbol{r}\right)\right]+F_{\text{ne}}\left[\phi\left(\boldsymbol{r}\right),\varepsilon_{ij}\left(\boldsymbol{r}\right)\right], \label{eq:SM_full_nemato-elastic_free_energy}
\end{equation}
where the bare electronic nematic and bare elastic free energies assume the form 
\begin{align}
    F_{\text{n}}\left[\phi\left(\boldsymbol{r}\right)\right]&=\frac{1}{2}\int_{r}\left\{ r\phi^{2}\left(\boldsymbol{r}\right)+c\left[\nabla\phi\left(\boldsymbol{r}\right)\right]^{2}\right\} +u\int_{r}\phi^{4}\left(\boldsymbol{r}\right), \label{eq:SM_bare_electronic_nematic_free_energy}
    \\
    F_{\text{e}}\left[\varepsilon_{ij}\left(\boldsymbol{r}\right)\right]&=\frac{1}{2}\int_{r}\left\{ B\left(\varepsilon_{xx}+\varepsilon_{yy}\right)^{2}+\mu_{1}\left(\varepsilon_{xx}-\varepsilon_{yy}\right)^{2}+\mu_{2}\left(2\varepsilon_{xy}\right)^{2}\right\}.\label{eq:SM_bare_elastic_free_energy}
\end{align}
The nemato-elastic bilinear, meanwhile, is given by 
\begin{equation}
    F_{\text{ne}}\left[\phi\left(\boldsymbol{r}\right),\varepsilon_{ij}\left(\boldsymbol{r}\right)\right]=-\lambda\int_{r}\phi\left(\boldsymbol{r}\right)[2\varepsilon_{xy}\left(\boldsymbol{r}\right)],
\end{equation}
In the equations above, we use the shorthand $\int_r \equiv \int \text{d}^2 r$. The quantity $r\propto T-T^*_0$ tunes through the bare nematic transition temperature, $T_0^*$,  $c$ is the bare nematic order parameter stiffness, and $u>0$ stabilizes the free energy. The bare elastic free energy is expanded to the harmonic level for the planar tetragonal system, where symmetry allows for three distinct elastic constants, $B$, $\mu_1$, and $\mu_2$, all of which must be positive for thermodynamic stability. The quantity $B$ is the in-plane bulk modulus, $\mu_1$ is the deviatoric modulus, and $\mu_2$ is the shear modulus. The bulk modulus characterizes the energy scale of symmetry-preserving, volume-changing strains, whereas the deviatoric and shear moduli account for the energy required for symmetry-breaking, volume-preserving strains. The quantity, $\lambda$, is the nemato-elastic coupling.

It is more convenient to express the elastic free energy in terms of the following three quantities:
\begin{align}
    \mu_{m}	\equiv\frac{\mu_{1}+\mu_{2}}{2}, && 
b	\equiv\frac{B}{\mu_{m}}, &&
\delta\mu	\equiv\frac{\mu_{1}-\mu_{2}}{\mu_{1}+\mu_{2}},
\end{align}
where $\mu_m$ is the typical energy scale of symmetry-lowering strains, $b$ is a non-universal dimensionless quantity relating the bulk and shear rigidities, and $\delta\mu$ is the dimensionless ``tetragonal anisotropy'' coefficient. The coefficient, $\delta\mu$, vanishes for two-dimensional hexagonal crystals and isotropic solids. With these quantities, the elastic free energy is 
\begin{equation}
    F_{\text{e}}\left[\varepsilon_{ij}\left(\boldsymbol{r}\right)\right]=\frac{\mu_{m}}{2}\int_{r}\left\{ b\left(\varepsilon_{xx}+\varepsilon_{yy}\right)^{2}+\left(1+\delta\mu\right)\left(\varepsilon_{xx}-\varepsilon_{yy}\right)^{2}+\left(1-\delta\mu\right)\left(2\varepsilon_{xy}\right)^{2}\right\}.
\end{equation}

Due to the Saint Venant Compatibility Relations discussed in the main text, and in nemato-elasticity problems within Refs. \citep{meeseConsequencesnematoelasticityStructurally2024, Meese_short_paper_2025, Meese_long_paper_2025}, the three strain components of a spatially inhomogeneous deformation are interdependent. This interdependence is the result of the relationship between the strain tensor---the physical quantity which costs elastic energy when nonzero---and the lattice displacement vector. These conditions are inviolate in ideal materials, and their violation in actual samples is a direct consequence of structural disorder in the form of crystalline defects   \citep{eshelbyContinuumTheoryLattice1956, dewitLinearTheoryStatic1970, dewitTheoryDisclinationsII1973, dewitTheoryDisclinationsIII1973, dewitTheoryDisclinationsIV1973, kronerekkehartContinuumTheoryDefects1981, muratoshioMicromechanicsDefectsSolids1987, kleinertGaugeFieldsSolids1989, grogerDefectinducedIncompatibilityElastic2008, beekmanDualGaugeField2017, pretkoFractonElasticityDuality2018, pretkoCrystaltofractonTensorGauge2019, gaaFractonelasticityDualityTwisted2021, meeseConsequencesnematoelasticityStructurally2024}.  Indeed, the fact that the linear strain tensor, $\varepsilon_{ij}$, follows from the displacement vector, $\boldsymbol{u}$, as 
\begin{equation}
    \varepsilon_{ij}\left(\boldsymbol{r}\right)=\frac{1}{2}\left[\partial_{i}u_{j}\left(\boldsymbol{r}\right)+\partial_{j}u_{i}\left(\boldsymbol{r}\right)\right],
\end{equation} 
implies there are only ever two independent strain fields since there are only two components of the displacement vector in two dimensions. Thus, the full nemato-elastic problem in 2D involves \textit{three} independent degrees of freedom: one Ising nematic order parameter, and two independent strain components. To proceed, it is convenient to use a ``helical basis'' for the strain fields that emphasizes the symmetry and intrinsic directionality of the compatibility relations \citep{Meese_short_paper_2025, Meese_long_paper_2025}. By doing so, one arrives at an elastic free energy in the form 
\begin{equation}
    F_{\text{e}}\left[\varepsilon_{ij}\left(\boldsymbol{r}\right)\right]=\frac{1}{2V}\sum_{\boldsymbol{q}}\boldsymbol{\varepsilon}_{h}^{\dagger}\cdot\mathfrak{C}\left(\hat{q}\right)\cdot\boldsymbol{\varepsilon}_{h}, \label{eq:bare_helical_elastic_free_energy}
\end{equation}
where $\boldsymbol{\varepsilon}_h \equiv (\varepsilon_1^h\; \varepsilon_2^h)^{\text{T}}$ is a vector of the Fourier amplitudes of the longitudinal ($\varepsilon^h_1$) and transverse strain components ($\varepsilon^h_2$). The terms ``longitudinal'' and ``transverse'' are defined in this context with respect to the wave vector, $\boldsymbol{q}$, and, following Refs. \cite{Meese_long_paper_2025, Meese_short_paper_2025}, are given in terms of the displacement vector, $\boldsymbol{u}$, as
\begin{equation}
    \begin{aligned}
        \varepsilon^h_1 &\equiv \text{i}\boldsymbol{q} \cdot \boldsymbol{u},
        \\
        \varepsilon^h_2 &\equiv \text{i}\hat{z}\cdot (\boldsymbol{q} \times \boldsymbol{u}).
    \end{aligned} \label{eq:helical_strains_from_displacements}
\end{equation}
The elastic stiffness matrix, written in the helical basis, is given by 
\begin{equation}
    \mathfrak{C}\left(\vec{q}\right)=\mu_{m}\left\{ b\boldsymbol{Q}_{A_{1g}}\boldsymbol{Q}_{A_{1g}}^{\text{T}}+\left(1+\delta\mu\right)\boldsymbol{Q}_{B_{1g}}\boldsymbol{Q}_{B_{1g}}^{\text{T}}+\left(1-\delta\mu\right)\boldsymbol{Q}_{B_{2g}}\boldsymbol{Q}_{B_{2g}}^{\text{T}}\right\} ,
\end{equation}
with the two-component, momentum-dependent form factors being given by 
\begin{equation}
    \begin{array}{ccc}
\boldsymbol{Q}_{A_{1g}}=\begin{pmatrix}1\\
0
\end{pmatrix}, & \boldsymbol{Q}_{B_{1g}}=\begin{pmatrix}\cos2\zeta\\
-\sin2\zeta
\end{pmatrix}, & \boldsymbol{Q}_{B_{2g}}=\begin{pmatrix}\sin2\zeta\\
\cos2\zeta
\end{pmatrix}.\end{array} \label{eq:SM_Qvectors}
\end{equation}
In the above, the momentum is defined in polar coordinates by $\boldsymbol{q}\equiv (q_x, q_y) \equiv q (\cos\zeta, \sin\zeta)$. These form factors transform the longitudinal, $\varepsilon_1^h$, and the transverse, $\varepsilon_2^h$, strains into the irreducible representations of the strain tensor of tetragonal point group, $\text{D}_{4h}$:
\begin{equation}
\begin{aligned}
    \varepsilon_{A_{1g}} &\equiv \varepsilon_{xx} + \varepsilon_{yy} = \boldsymbol{Q}_{A_{1g}}^{\text{T}} \cdot \boldsymbol{\varepsilon}_h   \\
    \varepsilon_{B_{1g}} &\equiv \varepsilon_{xx} - \varepsilon_{yy} = \boldsymbol{Q}_{B_{1g}}^{\text{T}} \cdot \boldsymbol{\varepsilon}_h 
    \\
    \varepsilon_{B_{2g}} &\equiv 2\varepsilon_{xy} = \boldsymbol{Q}_{B_{2g}}^{\text{T}} \cdot \boldsymbol{\varepsilon}_h.
\end{aligned}\label{eq:SM_strain_irreps_helical}
\end{equation}
Comparing Eqs. \eqref{eq:SM_Qvectors} and \eqref{eq:SM_strain_irreps_helical}, it is clear that the compatibility relations mix the symmetry-breaking strains, $\varepsilon_{B_{1g}}$ and $\varepsilon_{B_{2g}}$, with the symmetry-preserving dilatation strain $\varepsilon_{A_{1g}}= \varepsilon^h_1$. This will generally lead to non-local, long-ranged, coupling between the local electronic nematic order parameter and the local strain tensor. Generally, therefore, there is no direct proportionality between the \textit{local} nematic order parameter and the \textit{local} symmetry-breaking shear. This nonlocal relationship exists within bare elasticity theory, and even in the isotropic medium (see Sec.~\ref{supp:visualizing_strain_waves} and Refs.~\cite{Meese_short_paper_2025, Meese_long_paper_2025}). The nonlocality is present unless either $\cos2\zeta = 0$ or $\sin2\zeta = 0$, respectively. In the latter case, $\sin 2\zeta$ vanishes along the tetragonal coordinate axes, $[100]_T$ and $[010]_T$, we will be able to establish a local proportionality between the shear strain and the nematic order parameter.

One proceeds by minimizing the total free energy with respect to the two strain components, yielding the following coupled equations of state
\begin{equation}
    \mathfrak{C}\left(\hat{q}\right)\cdot\boldsymbol{\varepsilon}_{h}=\lambda\phi\boldsymbol{Q}_{B_{2g}}.
\end{equation}
Along the crystal axes, $\sin2\zeta = 0$ and $\cos2\zeta = \pm1$, in which case these equations of state decouple as 
\begin{equation}
    \begin{aligned}
        \mu_m(1+b+\delta\mu) \varepsilon^h_1 &= 0,
        \\
         2\mu_m(1-\delta\mu) \varepsilon_{xy} &= \lambda \phi,
    \end{aligned}
\end{equation}
since $2\varepsilon_{xy} = \pm \varepsilon^h_2$ in these directions. In real-space, these expressions yield that the shear strain, as a function of position, is given by $\varepsilon_{xy}(\boldsymbol{r}) \propto \phi(\boldsymbol{r})$. The fact that the direct proportionality exists between the induced shear strain and the electronic nematic order for these high-symmetry directions is a manifestation of ``direction-selective criticality'' known from ferroelastic \cite{cowleyAcousticPhononInstabilities1976, folkCriticalStaticsElastic1976} and electronic nematic \citep{karahasanovicElasticCouplingSpindriven2016, paulLatticeEffectsNematic2017, fernandesNematicityTwistRotational2020, heckerPhononinducedRotationElectronic2022} phase transitions, and is ultimately a result of the compatibility relations (SVCR) \citep{meeseConsequencesnematoelasticityStructurally2024, Meese_short_paper_2025, Meese_long_paper_2025}. Assuming, for simplicity, that spatial modulations occur  exclusively along $[100]_T$, then after substituting in the strain fields into the free energy, we obtain an effective one-dimensional nematic free energy of the form 
\begin{equation}
F_{\text{eff}}\left[\phi\left(\boldsymbol{r}\right)\right]=\frac{1}{2}\int_{x}\left\{ \left(r - \frac{\lambda^2}{\mu_2}\right)\phi^{2}\left(x\right)+c\left[\partial_x\phi\left(x\right)\right]^{2}\right\} +u\int_{x}\phi^{4}\left(x\right), \label{eq:SM_eff_electronic_nematic_free_energy}
\end{equation}
which is the free energy quoted in the main text when we define the energy units such that $c^2 = 1$ and set $r - \lambda^2/\mu_2 \equiv a(T - T^*)$, where $T^*$ is the effective transition temperature. One observes that for these modulations, the effective theory no longer depends on the non-universal parameter $b = 2B/(\mu_1+\mu_2)$. Because of the direct local proportionality between $\varepsilon_{xy}(x)$ and $\phi(x)$ along $[100]_T$ (and $[010]_T$), the spontaneous development of strain is therefore expressed through the spontaneous development of electronic nematic order which occurs when $r = \lambda^2 /\mu_2$.

\subsection{Partial square wave \textit{ansatz} as a model of twin formation}

As discussed in the main text, the mesoscale spatial modulations are observed within individual twin components. Since the boundaries of the twin components are not directly observable in the experiment, we have simplified the problem of twin formation to that of an over-constrained minimization of the following one-dimensional Ginzburg-Landau free energy:
\begin{equation}
    F_{\text{eff}}\left[\phi\left(x\right)\right]=\frac{1}{2}\int_0^L \text{d}x\,\left\{ a(T - T^* )\phi^{2}\left(x\right)+\left[\partial_x\phi\left(x\right)\right]^{2}\right\} +u\int_0^L \text{d}x\,\phi^{4}\left(x\right). \label{eq:SM_eff_electronic_nematic_free_energy_1D}
\end{equation}
The equation above appears in the main text, where the spatially modulated $B_{2g}$ electronic nematic order parameter is $\phi(x)$. The quantity $a(T - T^*)$ is the inverse (renormalized) nematic susceptibility with the nemato-elastic transition temperature being $T^*$ enhanced from $T_0^*<T^*$ by elastic fluctuations. The parameter $a>0$. The quartic coefficient, $u$, is strictly positive for thermodynamic stability. 

Given that the modulations occur along the $[100]_T$-direction, the SVCR decouple $\phi(x)$ from nonlocal dilatation strains, allowing for a local proportionality between the electronic nematic order parameter and the shear strain: $\phi(x) \propto \varepsilon_{xy}(x)$. While the development of the strain in this material is due to electronic nematicity, the \textit{local} proportionality between it and the spatially-modulated shear strain would allow for a similar free energy written entirely in terms of $\varepsilon_{xy}(x)$. This suggests a broader universality for strain waves applying beyond electronic nematicity in the iron pnictides. It is important, however, to recall for arbitrary modulations with off-axis momenta, that this universality would not exist.

\begin{figure}
    \centering
    \includegraphics[width=0.8\linewidth]{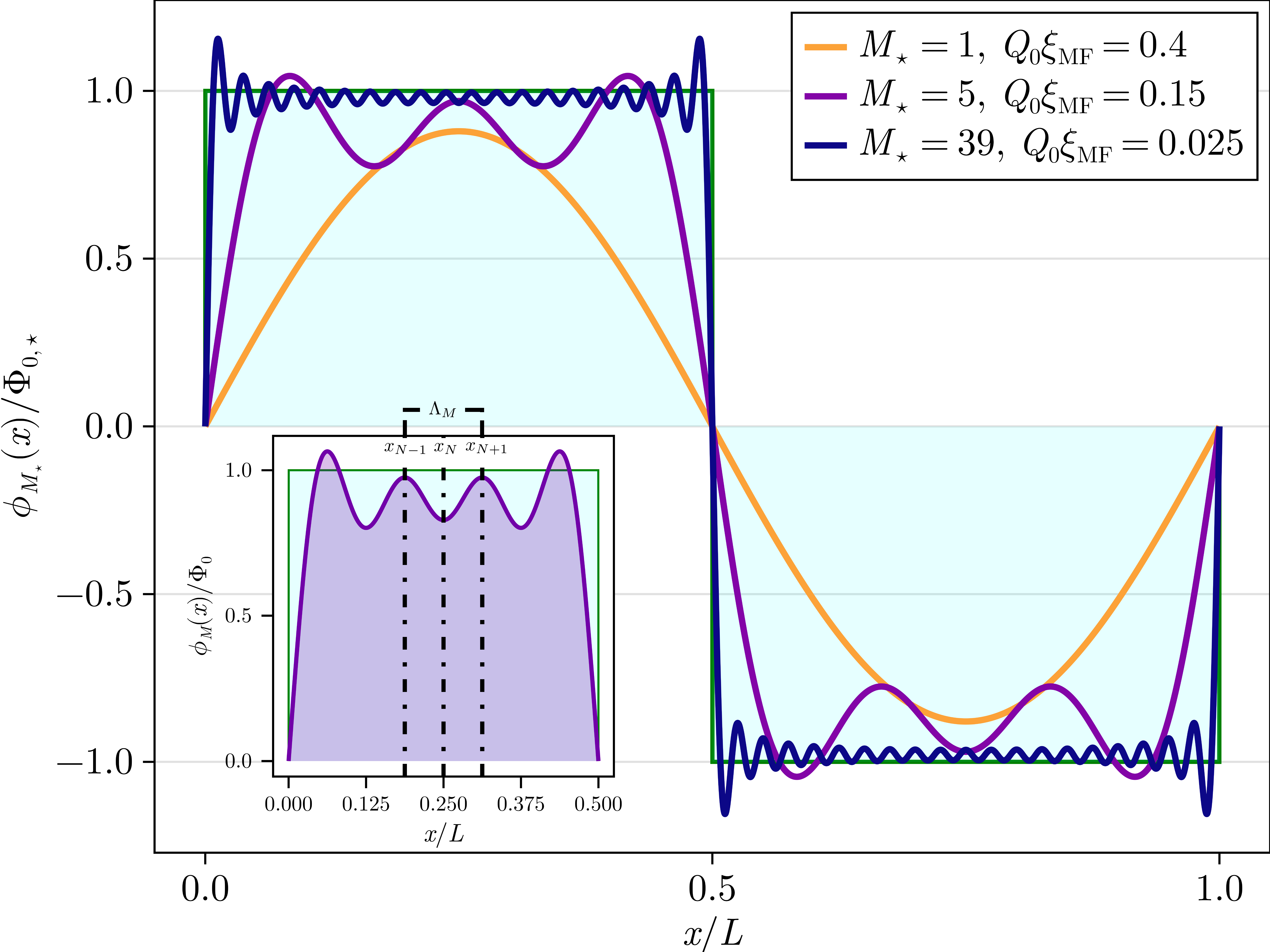}
    \caption{Twin formation through partial square waves as function of temperature. The minimizing twin domain field configuration, $\phi_{M_\star}(x)$,  is given for various temperatures, parameterized through the quantity $Q_0\xi_{\text{MF}} \propto 1/\sqrt{T^* - T}$ (see Eq. \eqref{eq:SM_Q0xi_proxy}). The field configuration is measured relative to $\Phi_{0,\star} = \sqrt{-r/4u}$ -- the order parameter associated with uniform long-range order. The perfectly uniform twin domains are shaded for comparison. As the system cools from the nematic critical temperature, $T^*$, spatial modulations within the bulk of each twin component appear. The length scale is given by $\Lambda_M$, as shown in the inset. The vertical lines denote particular extrema, as explained in the text, and it is found that $\Lambda_{M_\star} = L/(M_{\star} + 1)$.}
    \label{fig:SM_twin_formation}
\end{figure}

For twin formation to stabilize Eq. \eqref{eq:SM_eff_electronic_nematic_free_energy_1D} on the interval $x\in (0,L)$, one must constrain the nematic order parameter to be nonzero within either twin component, but take on opposite signs. This implies that $\phi(x = L/2) = 0$, if we assume the twin is symmetric. We are free to assume that $\phi(x \in (0, L/2)) > 0$, which implies that $\phi(x \in (L/2, L)) < 0$. However, elasticity requires that the traction on each boundary with unit normal $\boldsymbol{n}$ vanish such that $\sigma_{ij}n_j = 0$, where $\sigma_{ij}$ is the stress tensor \citep{landauTheoryElasticity1970, muratoshioMicromechanicsDefectsSolids1987}. In the absence of $\varepsilon_{xx}$ or $\varepsilon_{yy}$  uniaxial strains, the traction-free boundary condition reduces to a vanishing shear: $\varepsilon_{xy}(x = 0)=\varepsilon_{xy}(x = L)=0$. These constitute five boundary conditions for Eq. \eqref{eq:SM_eff_electronic_nematic_free_energy_1D}. However, minimizing the functional yields a one-dimensional nonlinear Helmholtz equation whose solution is only uniquely determined by \textit{two} boundary conditions, not five. Thus, the problem of twin formation is over-constrained, and we instead employ a variational approach. The \textit{ansatz} chosen is the partial square-wave, shown in the main text, and written as
\begin{equation}
    \phi_M(s) \equiv \Phi \, \mathcal{S}_M(s)  \equiv \Phi \cdot \frac{4}{\pi} \sum_{m \; \text{odd}}^M \frac{\sin(2\pi m s)}{m},\;s\equiv \frac{x}{L}\in [0,1], \label{eq:SM_partial_square_wave_ansatz}
\end{equation}
with cutoff integer $M$. This cutoff interpolates between a single sinuoidal waveform for $M = 1$ and perfectly flat twin components in the limit $M \rightarrow \infty$. In both extremes, there are no additional modulations within either twin component, however, they appear for all intermediate, finite values of $M \geq 3$, as shown in Fig.~\ref{fig:SM_twin_formation}. The amplitude $\Phi$ controls the magnitude of the uniform strain component in the twin.

Substituting in the \textit{ansatz} in Eq. \eqref{eq:SM_partial_square_wave_ansatz} yields a free energy density of the form
\begin{equation}
    f(\Phi, M) \equiv \frac{F[\phi_M(x)]}{L} = \frac{1}{2}|r| \tilde{r}(M)\Phi^2 + u \tilde{u}(M)\Phi^4,
\end{equation}
which depends only on the variational parameters, $\Phi$, and $M$. The quantities, $\tilde{r}(M)$ and $\tilde{u}(M)$ are determined by 
\begin{equation}
    \begin{aligned}
        \tilde{r}(M) &\equiv \int_0^1 \text{d}s\,\left\{ \text{sgn}(r) \mathcal{S}_M^2(s) + (Q_0\xi_{\text{MF}})^2\left[ \frac{\partial_s\,\mathcal{S}_M(s)}{2\pi} \right]^2  \right\},
        \\
        \tilde{u}(M) &\equiv \int_0^1 \text{d}s\, \mathcal{S}_M^4(s),
     \end{aligned}
\end{equation}
where $\xi_{\text{MF}} \equiv 1/\sqrt{a|T - T^*|}$ and $Q_0 \equiv 2\pi/L$ is the fundamental harmonic. In the usual case of the $\Phi^4$-model with a uniform order parameter---one that cannot satisfy the requisite boundary conditions for twin formation---one replaces $\tilde{r}(M)$ and $\tilde{u}(M)$ with $\text{sgn}(r)$ and $1$, respectively. The parameter, $Q_0\xi_{\text{MF}}$, is written explicitly as 
\begin{equation}
    Q_0\xi_{\text{MF}} = \frac{2\pi\xi_{\text{MF}}}{L} = \frac{2\pi}{L\sqrt{a|T -T^*|}}, \label{eq:SM_Q0xi_proxy}
\end{equation}
and parametrizes the temperature $T$ in the ordered phase. 

Minimizing with respect to the amplitude $\Phi$ yields 
\begin{equation}
    \begin{aligned}
        \Phi_\star^2 &= -\frac{|r| \tilde{r}(M)}{4u \tilde{u}(M)},
        \\
        f(\Phi_\star,M) &= \begin{cases}
            0, &r \propto T - T^* >0
            \\
            -\frac{r^2 \tilde{r}^2(M)}{16 u \tilde{u}(M)}, & r \propto T - T^* <0 
        \end{cases}
    \end{aligned}
\end{equation}
Minimizing field configurations, $\phi_{M_\star}(x)$, are shown in Fig.~\ref{fig:SM_twin_formation} for various temperatures, parameterized through the quantity $Q_0\xi_{\text{MF}} = 2\pi \xi_{\text{MF}}/L \propto 1/\sqrt{T^* - T}$.

\begin{figure}
    \centering
    \includegraphics[width=0.8\linewidth]{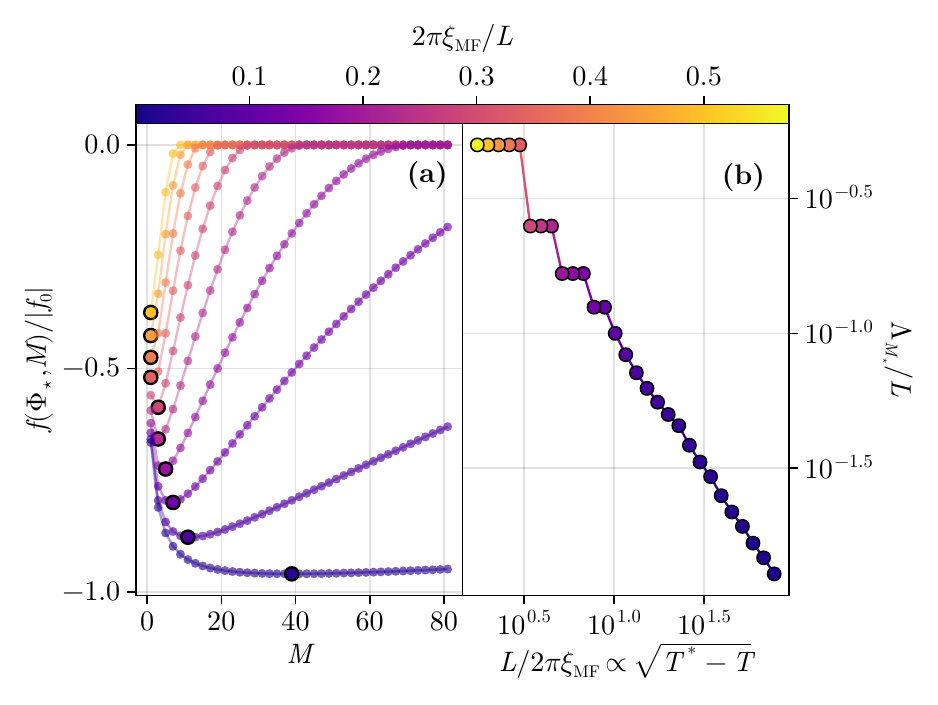}
    \caption{\textit{Ansatz} minimizers as a function within the nematic phase $(T < T^*)$. \textbf{(a)} The free energy density $f(\Phi_\star, M)$ as a function of the cutoff integer $M$. The free energy is evaluated at the minimizing amplitude $\Phi_\star$, and is measured relative to the free energy minimum with a uniform nematic order parameter, $f_0$. The different curves correspond to different temperatures, $T$, from Eq. \eqref{eq:SM_Q0xi_proxy}. The minimizing cutoff, $M_\star$, is emphasized with a larger outlined point. \textbf{(b)} The modulation length within the bulk of a twin component, $\Lambda_{M_\star}$, as a function of the distance from nematic criticality. In both panels, the lines are guides to the eye.}
    \label{fig:SM_free_energy_minima}
\end{figure}

From the expressions above, it is clear that the partial square wave \textit{ansatz} is only nonzero in the nematic phase with $T < T^*$. Comparing with the minimizers for a uniform order parameter, one can write the conditions above more succinctly as
\begin{equation}
    \begin{aligned}
        \Phi^2_{\star} &= \text{sgn}(r)\left[ \frac{\tilde{r}(M_\star)}{\tilde{u}(M_\star)}\right] \Phi_{0, \star}^2,
        \\
        f(\Phi_\star,M) &= - \frac{\tilde{r}^2(M)}{\tilde{u}(M)}\, \vert f_0 \vert,
    \end{aligned}
\end{equation}
where $\Phi_{0,\star}^2 \equiv -r/4u$ is the minimizing amplitude for the uniform case, and $f_0$ is the corresponding minimum free energy density. For $T > T^*$, $f_0 = 0$ and for $T < T^*$, $f_0 = -r^2/16u$.  By varying the odd integer $M$ at various temperatures, one observes a minimum in the free energy develop for $M > 1$, with the minimizing value of $M$, denoted as $M_\star$, as shown in Fig.~\ref{fig:SM_free_energy_minima}(a). This shows that uniformity within either twin component is unstable towards additional spatial modulations on top of a bulk background. 

We quantify the length scale associated with these spatial modulations through a quantity, $\Lambda_M$, which is the distance between the extrema immediately adjacent to the center of either twin component, as shown in the inset of Fig.~\ref{fig:SM_twin_formation}. One can obtain a closed form expression for $\partial_s\,\mathcal{S}_M(s)$ by summing a partial geometric series to obtain
\begin{equation}
    \partial_s\, \mathcal{S}_M(s) = \partial_s S_M(s) = \frac{4}{\pi} \cdot 2\pi \left\{ \frac{\sin\left[2\pi(M+1)s \right]}{2\sin(2\pi s)} \right\},
\end{equation}
from which it follows that the bulk extrema occur when $2\pi(M+1)x/L = p\pi$ for $p\in \mathds{Z}^+$. As shown in the inset of Fig.~\ref{fig:SM_twin_formation}, the center of the twin domain is $x/L = 1/4$, which then corresponds to the integer $N \equiv (M + 1) / 2$. Since $M$ is odd, then $N$ represents the number of terms included in $\mathcal{S}_M(s)$. In the inset, we mark $x_N \equiv L/4$. The adjacent extrema therefore correspond to $x_{N\pm 1}$. The spatial modulation scale, $\Lambda_M$ follows as
\begin{equation}
    \Lambda_M \equiv x_{N+1} - x_{N-1} = L \left\{ \frac{N + 1}{2(M+1)} -\frac{N - 1}{2(M+1)} \right\} = \frac{L}{M+1}.
\end{equation}
The value of this length scale evaluated for the minimizing cutoff, $\Lambda_{M_\star}$, is shown in Fig.~\ref{fig:SM_free_energy_minima}(b), as a function of $1/Q_0\xi_{\text{MF}} = L/2\pi\xi_{\text{MF}}$. As the temperature decreases from nematic criticality, one finds that $M_\star \sim L/\xi_{\text{MF}} \propto \sqrt{T^* - T}$, showing that the modulation scale, $\Lambda_{M_\star}$, is a slowly decreasing function of the temperature: $\Lambda_{M_\star} \sim 1/\sqrt{T^* - T}$. 

\begin{figure}
    \centering
    \includegraphics[width=0.8\linewidth]{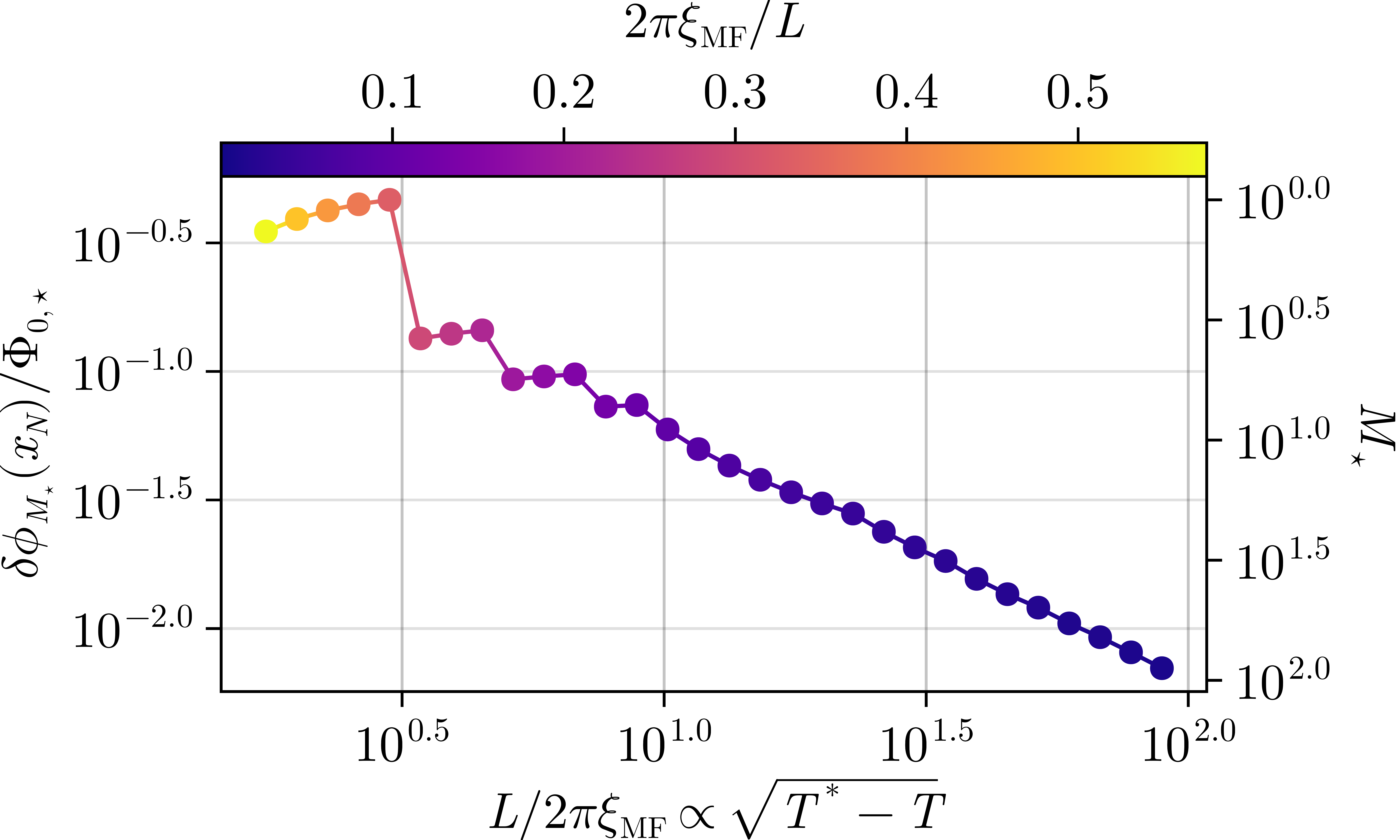}
    \caption{Spatial modulation amplitude within the bulk of the twin domains as a function of temperature. The temperature scale is parameterized through Eq. \eqref{eq:SM_Q0xi_proxy}, and it shows that the bulk modulation amplitude is a slowly decreasing function of temperature deep in the nematic phase. The right descending axis shows the corresponding minimizing cutoff, $M_\star$, in the partial square wave \textit{ansatz}. All axes are on a logarithmic scale. The line is a guide for the eye.}
    \label{fig:SM_bulk_modulation_amplitude}
\end{figure}

Similarly, by evaluating the \textit{ansatz} at $x_{N+1}$ and $x_N$, one can determine the amplitude of the spatial modulations within the twin domains as
\begin{equation}
    \delta\phi_{M_\star}(x_N) \equiv \frac{1}{2}\left\vert \phi_{M_\star}(x_{N+1}) - \phi_{M_\star}(x_{N}) \right\vert.
\end{equation}
This bulk modulation amplitude is plotted in Fig.~\ref{fig:SM_bulk_modulation_amplitude} as a function of temperature. Far from nematic criticality, it is observed that the amplitude is also a slowly decreasing function of the temperature as well, $\delta \phi_{M_\star} \sim 1/\sqrt{T^* - T}$, just as the bulk modulation scale, $\Lambda_{M_\star}$.

\section{Visualizing elastic strain waves\label{supp:visualizing_strain_waves}}

In this section, we demonstrate how to compute the two-dimensional displacement vector, $\boldsymbol{u}(\boldsymbol{r})$, from a known spatially modulated strain field. In doing so, the strain wave necessarily satisfies the Saint-Venant compatibility relations, and from the displacement vector one can compute the three components of the strain tensor: the dilatation strain $\varepsilon_{xx} + \varepsilon_{yy}$, the deviatoric strain $\varepsilon_{xx} - \varepsilon_{yy}$, and the shear strain $\varepsilon_{xy}$. We will show that generally there is a nonlocal relationship between the spatially modulated strain tensor strictly enforced by the compatibility relations\cite{Meese_long_paper_2025, Meese_short_paper_2025}. However, we show that if the modulations have specific momentum directions, then the nonlocal relationship collapses to a local direct proportionality. We will then simplify the analysis to the case of strain waves with a single well-defined wave vector, and use these results to create Fig.~\ref{fig:shear_waves} in the main text. To keep the discussion regarding the geometry universal, we refrain in this section from discussing the mechanism for a strain wave, instead focusing on the impact the resulting deformation has on a medium  should the wave exist.

We start by assuming the helical strain amplitudes from Eq. \eqref{eq:helical_strains_from_displacements} are known functions of the wave vector $\boldsymbol{q}$, $\boldsymbol{\varepsilon}_h = \boldsymbol{\varepsilon}_h(\boldsymbol{q})$. We will simplify down to a single static shear wave after obtaining the general solution. By inverting Eq. \eqref{eq:helical_strains_from_displacements}, it follows that the displacement vector amplitudes are 
\begin{equation}
    \begin{bmatrix}
        u_x(\boldsymbol{q}) \\ u_y(\boldsymbol{q})
    \end{bmatrix} = -\frac{\text{i}}{q} \begin{bmatrix}
        \cos(\zeta) & -\sin(\zeta)
        \\
        \sin(\zeta) & \cos(\zeta)
    \end{bmatrix} \begin{bmatrix}
        \varepsilon_1^h(\boldsymbol{q}) \\ \varepsilon_2^h(\boldsymbol{q})
    \end{bmatrix} \equiv -\frac{\text{i}}{q} \mathcal{R}(\hat{q}) \cdot \boldsymbol{\varepsilon}_h(\boldsymbol{q}),
\end{equation}
with the wave vector being parameterized by $\boldsymbol{q}\equiv q(\cos(\zeta),\sin(\zeta))$. Inverting the Fourier transform, one obtains the general solution for displacement vector in real-space as 
\begin{equation}
    \boldsymbol{u}(\boldsymbol{r}) = -\begin{bmatrix}
        \partial_x & -\partial_y
        \\
        \partial_y & \partial_x
    \end{bmatrix} \cdot \boldsymbol{W}(\boldsymbol{r}), \label{eq:general_solution_displacement_vector_real-space}
\end{equation}
where the vector, $\boldsymbol{W}(\boldsymbol{r})$, is defined by 
\begin{equation}
    \boldsymbol{W}(\boldsymbol{r}) \equiv \int \text{d}^2 r^\prime \, g(\boldsymbol{r} - \boldsymbol{r}^\prime) \boldsymbol{\varepsilon}_h(\boldsymbol{r}^\prime),
\end{equation}
and corresponds to the nonlocal propagation of helical elastic strain throughout the medium. In calculating Eq.~\eqref{eq:general_solution_displacement_vector_real-space}, we have used $\boldsymbol{\varepsilon}_h(\boldsymbol{r}) \equiv \frac{1}{V}\sum_{\boldsymbol{q}} \text{e}^{\text{i}\boldsymbol{q}\cdot \boldsymbol{r}}\boldsymbol{\varepsilon}_h(\boldsymbol{q})$, and in $\boldsymbol{W}$, the function $g(\boldsymbol{r})$ is the Green's function for the Poisson equation in 2D: $g(\boldsymbol{r}) = -\frac{1}{2\pi}\log(r)$ \cite{Meese_long_paper_2025}. This integration is based on the compatibility relations using the helical strain formalism, and produces the lattice displacement vector $\boldsymbol{u}(\boldsymbol{r})$ from a compatible helical strain field uniquely, up to a global \textit{uniform} translation and rotation. Each unit cell,  represented schematically in Fig.~\ref{fig:shear_waves} of the main text as a circle and initially positioned at $\boldsymbol{r}$, is displaced according to the deformation: $\boldsymbol{r}\rightarrow\boldsymbol{r} + \boldsymbol{u}(\boldsymbol{r})$. This displacement vector, being the integrated \textit{strain}, rather than integrated \textit{stress}, is the solution in any crystal with any point group rotational symmetry.

From Eq. \eqref{eq:general_solution_displacement_vector_real-space}, the dilatation and shear strains can be computed as follows
\begin{equation}
    \begin{aligned}
        \varepsilon_{A_{1g}}(\boldsymbol{r}) &= \boldsymbol{\nabla} \cdot \boldsymbol{u}(\boldsymbol{r}) = \varepsilon^h_1(\boldsymbol{r}),
        \\
        \varepsilon_{x^2-y^2}(\boldsymbol{r}) &= \partial_x u_x(\boldsymbol{r}) - \partial_y u_y(\boldsymbol{r}) = -\left(\partial_x^2 - \partial_y^2\right) W_1(\boldsymbol{r})+\left(2\partial_x\partial_y\right)W_2(\boldsymbol{r}),
        \\
        \varepsilon_{xy}(\boldsymbol{r}) &= \frac{1}{2} \left[ \partial_x u_y(\boldsymbol{r}) + \partial_y u_x(\boldsymbol{r}) \right]= -\left(2\partial_x\partial_y\right) W_1(\boldsymbol{r}) -\left(\partial_x^2 -\partial_y^2\right)W_2(\boldsymbol{r}), 
    \end{aligned}\label{eq:general_solution_dilatation_shear_real-space}
\end{equation}
 These are general expressions for any 2D crystalline point group even though we have adopted the $A_{1g}$ label for the dilatation from the tetragonal point group $\text{D}_{4h}$. In deriving the dilatation above, we exploited the Green's function property that $\nabla^2 g(\boldsymbol{r}) = -\delta(\boldsymbol{r})$. The second and third lines emphasize the generally nonlocal interdependence between the dilatation, deviatoric, and shear strain components that exists for any crystal symmetry -- and persists even in the isotropic continuum \cite{Meese_short_paper_2025, Meese_long_paper_2025}. This nonlocal interdependence is present for generic spatial modulations, \textit{e.g.} when both partial derivatives are nonzero. To show this interdependence explicitly, we apply to the second (third) line of Eq.~\eqref{eq:general_solution_dilatation_shear_real-space} the operator $\partial_x^2 - \partial_y^2$ ($4\partial_x^2\partial_y^2$), and sum the results to find
 \begin{align}
     (\partial_x^2 -\partial_y^2)\varepsilon_{x^2-y^2}(\boldsymbol{r}) + (2\partial_x\partial_y)(2\varepsilon_{xy}(\boldsymbol{r})) &= -
     \left[(\partial_x^2 - \partial_x^2)^2 + (2\partial_x\partial_y)^2\right] W_1(\boldsymbol{r}) \nonumber\\
     &= -\nabla^4W_1(\boldsymbol{r}).
 \end{align}
 Using $\nabla^2g(\boldsymbol{r}) = -\delta(\boldsymbol{r})$ again, it follows that $\nabla^2W_1(\boldsymbol{r}) = -\varepsilon_1^h(\boldsymbol{r})$. Thus, we recover
 \begin{align}
     (\partial_x^2 -\partial_y^2)\varepsilon_{x^2-y^2}(\boldsymbol{r}) + (2\partial_x\partial_y)(2\varepsilon_{xy}(\boldsymbol{r})) = (\partial_x^2 +\partial_y^2)\varepsilon_{A_{1g}}(\boldsymbol{r}), \label{eq:supp_SVCR}
 \end{align}
 the exact Saint Venant compatibility relation quoted in the main text. Relabeling $\boldsymbol{r}$ as $\boldsymbol{r}^\prime$, multiplying both sides by $g(\boldsymbol{r} - \boldsymbol{r}^\prime)$, and integrating $\boldsymbol{r}^\prime$ over the infinite volume, we recover
 \begin{equation}
     \varepsilon_{A_{1g}}(\boldsymbol{r}) = -(\partial_x^2 -\partial_y^2)\int\text{d}^2 r^\prime\, g(\boldsymbol{r} - \boldsymbol{r}^\prime)\varepsilon_{x^2 - y^2}(\boldsymbol{r}^\prime) - (2\partial_x\partial_y)\int\text{d}^2 r^\prime\, g(\boldsymbol{r} - \boldsymbol{r}^\prime)[2\varepsilon_{xy}(\boldsymbol{r}^\prime)].
 \end{equation}
 To obtain the above result, we integrated by parts assuming that the strains vanish at infinity, and have used the derivative identity $\partial_{j^\prime}g(\boldsymbol{r} - \boldsymbol{r}^\prime) = -\partial_{j}g(\boldsymbol{r} - \boldsymbol{r}^\prime)$ to bring the partial derivatives outside of the integrals.

The manipulations above show that there is a nonlocal, modulation-direction-dependent, relationship between the three components of the strain tensor in 2D. If we assert that there is only one symmetry-breaking strain wave in the system, and choose it as the shear wave over the deviatoric wave, then the dilatation strain reduces to
\begin{equation}
    \varepsilon_{A_{1g}}(\boldsymbol{r}) =  - (2\partial_x\partial_y)\int\text{d}^2 r^\prime\, g(\boldsymbol{r} - \boldsymbol{r}^\prime)[2\varepsilon_{xy}(\boldsymbol{r}^\prime)], \label{eq:dilatation_nonlocal_shear}
\end{equation}
and establishes that the degree of dilatation strain is controlled by partial differentiation -- thus the direction of spatial modulations control the interdependence between the dilatation and the shear strains. If there is a shear wave that only depends on the $\hat{x} = [100]$ direction, namely $\partial_y\varepsilon_{xy}(\boldsymbol{r})=0$ and $\partial_x\varepsilon_{xy}(\boldsymbol{r})\neq0$, then it follows from the Green's function that the convolution only depends on $x$ as well. After partial differentiation, the dilatation vanishes, showing that the shear wave can grow arbitrarily large without inducing dilatation strains. Alternatively, taking the modulations in the shear wave to be only along the $[110]$-direction, such that $(\partial_x - \partial_y) \varepsilon_{xy}(\boldsymbol{r}) = 0$ while $(\partial_x + \partial_y) \varepsilon_{xy}(\boldsymbol{r}) \neq 0$, then it can be shown from Eq. \eqref{eq:dilatation_nonlocal_shear} that the dilatation and shear strains are directly proportional: $\varepsilon_{A_{1g}}(x+y) \propto \varepsilon_{xy}(x+y)$.

We now focus on the case of static waves with well-defined wave vector. Returning to the helical strain basis as the independent degrees of freedom, we seek the displacement vector and the three strain components that the helical strain waves induce. Assume that the helical strain is a static wave in real-space, such that 
\begin{equation}
    \boldsymbol{\varepsilon}_h(\boldsymbol{r}) \equiv \boldsymbol{\varepsilon}_0 \cos(\boldsymbol{Q}\cdot \boldsymbol{r}),
\end{equation}
where $\boldsymbol{\varepsilon}_0$ is a constant two-component vector. The helical strain amplitude follows from Fourier transformation as 
\begin{equation}
    \boldsymbol{\varepsilon}_h(\boldsymbol{q}) = \frac{1}{2} V\boldsymbol{\varepsilon}_0\,\left(\delta_{\boldsymbol{q}, \boldsymbol{Q}} + \delta_{\boldsymbol{q}, -\boldsymbol{Q}} \right),
\end{equation}
where $V \equiv L_x L_y$ is the macroscopic area of the 2D system. Substituting the above into Eq. \eqref{eq:general_solution_displacement_vector_real-space} yields 
\begin{equation}
    \boldsymbol{u}(\boldsymbol{r}) = \frac{1}{2\text{i}Q} \left[ \text{e}^{\text{i} \boldsymbol{Q} \cdot \boldsymbol{r}} \mathcal{R}(\hat{Q}) + \text{e}^{-\text{i} \boldsymbol{Q} \cdot \boldsymbol{r}} \mathcal{R}(-\hat{Q}) \right] \cdot \boldsymbol{\varepsilon}_0 = \frac{\sin(\boldsymbol{Q}\cdot\boldsymbol{r})}{Q} \begin{bmatrix}
        \varepsilon_{0,1}\cos(\theta_Q)  -\varepsilon_{0,2}\sin(\theta_Q)
        \\
        \varepsilon_{0,1}\sin(\theta_Q) + \varepsilon_{0,2}\cos(\theta_Q)
    \end{bmatrix}. \label{eq:displacement_vector_Q}
\end{equation}
where $\boldsymbol{Q} \equiv Q(\cos(\theta_Q),\sin(\theta_Q))$ and $\mathcal{R}(-\hat{Q}) = -\mathcal{R}(\hat{Q})$. The dilatation, deviatoric, and shear waves follow as 
\begin{equation}
    \begin{aligned}
        \varepsilon_{A_{1g}}(\boldsymbol{r}) &= \varepsilon_{0,1}\cos(\boldsymbol{Q} \cdot \boldsymbol{r}),
        \\
        \varepsilon_{x^2 - y^2}(\boldsymbol{r}) &= \left[\varepsilon_{0,1}\cos(2\theta_Q) - \varepsilon_{0,2}\sin(2\theta_Q)\right]\cos(\boldsymbol{Q} \cdot \boldsymbol{r}),
        \\
        \varepsilon_{xy}(\boldsymbol{r}) &= \frac{1}{2}\left[ \varepsilon_{0,1}\sin(2\theta_Q) + \varepsilon_{0,2}\cos(2\theta_Q) \right]\cos(\boldsymbol{Q} \cdot \boldsymbol{r}).
    \end{aligned}\label{eq:strain_components_helical_wave}
\end{equation}
When there is only one symmetry-breaking strain wave in system, the expressions above further simplifies. Consider the case  that between the deviatoric and shear strains, only the shear strain is nonzero. Then it follows that 
\begin{equation}
    \varepsilon_{x^2 - y^2}(\boldsymbol{r}) = 0\quad \Rightarrow \quad \varepsilon_{0,1}\cos(2\theta_Q) = \varepsilon_{0,2}\sin(2\theta_Q).\label{eq:no_deviatoric}
\end{equation}
We see from the above that if $\sin(2\theta_Q) = 0$, then the dilatation amplitude, $\varepsilon_{0,1}$ must vanish. This happens when the wave vector lies along the $[100]$ and $[010]$ axes. Thus, not only is the shear strain independent of the dilatation, without the deviatoric strain, the dilatation strain is exactly zero. Likewise, if $\cos(2\theta_Q) = 0$, then the transverse amplitude $\varepsilon_{0,2}$ must vanish. In this situation, which occurs along the $[110]$ and $[1\bar{1}0]$ axes, the opposite is true. Now the shear wave generates a displacement vector that maximizes the magnitude of the dilatation strain. For any other direction of momentum, it follows that $\varepsilon_{0,2} = \varepsilon_{0,1}\cot(2\theta_Q)$. Substituting this into the expression for the shear strain and simplifying yields
\begin{equation}
    \varepsilon_{A_{1g}}(\boldsymbol{r}) = \left[\sin(2\theta_Q)\right]\,2\varepsilon_{xy}(\boldsymbol{r}),
\end{equation}
showing that the amount of dilatation strain accompanying a shear wave is indeed directly controlled by the momentum direction, $\theta_Q$. This expression is equivalent to the Saint Venant compatibility relation in Eq.~\eqref{eq:dilatation_nonlocal_shear}, applied to the case of a shear wave with vanishing deviatoric strain. In the limit that the wave vector lies along the $[100]$ or $[010]$ axes, then $\varepsilon_{A_{1g}}(\boldsymbol{r}) =0$, regardless of the amplitude associated with the shear strain wave.

Fig.~\ref{fig:shear_waves} of the main text illustrates how the degree of interdependence of the shear and the dilatation waves is controlled by the modulation direction, $\theta_Q$. The former is softened near the tetragonal-to-orthorhombic phase transition, whereas the latter is not since the symmetry-preserving dilatation remains gapped at the transition \cite{cowleyAcousticPhononInstabilities1976, karahasanovicElasticCouplingSpindriven2016,Meese_short_paper_2025, Meese_long_paper_2025}. To produce the figure, we use the wave vector direction in Eq.~\eqref{eq:strain_components_helical_wave} and Eq.~\eqref{eq:no_deviatoric} to isolate the shear wave from the dilatation wave in Fig.~\ref{fig:shear_waves}\textbf{(a,c)} with $\theta_Q = 0$, and maximize the dilatation in Fig.~\ref{fig:shear_waves}\textbf{(b,d)} with $\theta_Q = \pi/4$. We then use Eq. \eqref{eq:displacement_vector_Q} to displace each unit cell by the appropriate local displacement vector. Starting with  $\theta_Q =0$, it follows that $\sin(2\theta_Q) = 0$, and Eq.~\eqref{eq:no_deviatoric} restricts the only nonzero solution for the strain wave to have helical amplitudes $\boldsymbol{\varepsilon}_0\equiv (0, \varepsilon_0)^{\text{T}}$. The dilatation and shear strains, as well as the displacement vector, are then
\begin{equation}
    \begin{aligned}
        \varepsilon_{A_{1g}}(x,y) &= 0, & u_x(x,y) &= 0, 
        \\
        \varepsilon_{xy} (x,y) &= \frac{1}{2}\varepsilon_{0} \cos(Qx),& u_y(x,y) &= \frac{\varepsilon_{0}}{Q}\sin(Qx),
    \end{aligned}
\end{equation}
which correspond to Fig.~\ref{fig:shear_waves}\textbf{(a,c)}. This shows that for $\hat{Q} \propto[100]$, the symmetry-breaking shear strain amplitude, $\varepsilon_0$, can grow arbitrarily large without inducing costly volume-changing dilatation. This is not generally true. For example, changing the direction to $\theta_Q =\pi/4$ such that $\cos(2\theta_Q) =0$, then Eq.~\eqref{eq:no_deviatoric} now restricts $\boldsymbol{\varepsilon}_0 =(\varepsilon_0,0)^{\text{T}}$. The strains and displacement vector then follow as
\begin{equation}
    \begin{aligned}
        \varepsilon_{A_{1g}}(x,y) &= \varepsilon_{0}\cos\left[\frac{Q(x + y)}{\sqrt{2}}\right], & u_x(x,y) &= \frac{\varepsilon_{0}}{Q\sqrt{2}}\sin\left[\frac{Q(x + y)}{\sqrt{2}}\right],
        \\
        \varepsilon_{xy} (x,y) &= \frac{1}{2}\varepsilon_{0}\cos\left[\frac{Q(x + y)}{\sqrt{2}}\right], & u_y(x,y) &= \frac{\varepsilon_{0}}{Q\sqrt{2}}\sin\left[\frac{Q(x + y)}{\sqrt{2}}\right],
    \end{aligned}
\end{equation}
These equations correspond to the strain and displacement wave in Fig.~\ref{fig:shear_waves}\textbf{(b,d)}. Clearly a compatible shear wave of any amplitude, $\varepsilon_0$, induces a simultaneous dilatation wave which incurs a higher energy cost if its wave vector satisfies $\cos(2\theta_Q) =0$.

\end{document}